\documentclass[fleqn,usenatbib]{mnras}

\usepackage{newtxtext,newtxmath}

\usepackage[T1]{fontenc}

\DeclareRobustCommand{\VAN}[3]{#2}
\let\VANthebibliography\thebibliography
\def\thebibliography{\DeclareRobustCommand{\VAN}[3]{##3}\VANthebibliography}

\usepackage{graphicx}	
\usepackage{subcaption} 
\usepackage{amsmath}	
\usepackage{color}
\usepackage{rotating}
\usepackage{array} 
\usepackage{booktabs}
\usepackage{pdflscape}
\usepackage{lastpage}
\usepackage[export]{adjustbox}
\usepackage{hyperref}
\usepackage{CJKutf8}
\usepackage[normalem]{ulem}

\defcitealias{yue_novel_2024}{Y24}
\title[Halo-jet link of radio quasar]{Revealing the link between halo mass and radio jet activities in quasars}

\author[B.-H. Yue et al.]{
B.-H. Yue~\begin{CJK*}{UTF8}{gbsn}(岳博涵)\end{CJK*}$^{1,2}$,\thanks{E-mail: bohan.yue@ed.ac.uk (BY)}
P. N. Best$^{1}$,
H. J. A. R\"ottgering$^{2}$,
K. J. Duncan$^{1}$,
C. L. Hale$^{1,3}$,
L. K. Morabito$^{4,5}$,
\newauthor
and D. J. B. Smith$^{6}$
\\
$^{1}$Institute for Astronomy, University of Edinburgh, Edinburgh EH9 3HJ, UK\\
$^{2}$Leiden Observatory, Leiden University, PO Box 9513, NL-2300 RA Leiden, The Netherlands\\
$^{3}$Astrophysics, Department of Physics, University of Oxford, Keble Road, Oxford, OX1 3RH, UK\\
$^{4}$Centre for Extragalactic Astronomy, Department of Physics, Durham University, Durham DH1 3LE, UK\\
$^{5}$Institute for Computational Cosmology, Department of Physics, University of Durham, South Road, Durham DH1 3LE, UK\\
$^{6}$Centre for Astrophysics Research, University of Hertfordshire, College Lane, Hatfield AL10 9AB, UK\\
}

\date{Accepted XXX. Received YYY; in original form ZZZ}

\pubyear{2026}

\begin{document}
\label{firstpage}
\pagerange{\pageref{firstpage}--\pageref{lastpage}}
\maketitle

\begin{abstract}

\noindent There is a fundamental lack of understanding as to why quasars that are otherwise very similar can have such a wide range of radio jet powers, and the large-scale environment is thought to play an important role. We investigate the spatial clustering properties of 225,382 quasars from the extended Baryon Oscillation Spectroscopic Survey (eBOSS) within the LOFAR Two-metre Sky Survey (LoTSS) Data Release 2 footprint, split by the statistically-calculated fraction of their radio flux densities contributed by jets (relative to the contribution from star formation). We find a positive correlation between the clustering strengths of quasars and their jet fraction, where quasars with a higher jet fraction have a higher clustering amplitude measured by their two-point correlation functions. We show that this correlation is unlikely related to differences in BH masses or bolometric luminosities. Quasars dominated by powerful jet activities generally reside in dark matter haloes $10-100$ times more massive than those without strong jets, with typical halo masses of $10^{13-14}\ h^{-1}M_\odot$, establishing a robust link between powerful AGN jets and rich cluster environments. Our results demonstrate that halo mass is important for determining the power of radio jets, but suggest that there is no minimum dark matter halo mass or BH mass required for the triggering of jets. The observed correlation suggests that BH spin is likely to play a minor role in jet production; instead, the key driver could be the presence of a strong magnetic flux.

\end{abstract}




\section{Introduction}
\label{sec:intro}

Quasars (QSOs) are luminous active galaxy nuclei (AGN) powered by actively accreting supermassive black holes (SMBHs). Despite being luminous in optical/infrared bands, only a small percentage ($5\sim10\%$) of quasars have been detected in previous large radio surveys, for example, the Faint Images of the Radio Sky at Twenty centimeters survey \citep[FIRST;][]{becker_first_1995}. These quasars are historically believed to host powerful relativistic jets and are referred to as the `radio-loud' (RL) quasar population \citep[e.g.][]{kellermann_vla_1989}, and their radio-undetected counterparts are called `radio-quiet' (RQ). However, as recent radio surveys, for example those in low-frequency radio bands carried out by the LOw-Frequency ARray \citep[LOFAR;][]{van_haarlem_lofar_2013}, probe into the radio sky to an order of magnitude deeper, it is clear that the distribution of quasar radio emission is continuous \citep[e.g.][]{balokovic_disclosing_2012,gurkan_lotsshetdex_2019,macfarlane_radio_2021,calistro_rivera_ubiquitous_2024,yue_novel_2024}. This is supported by direct radio observations of nearby galaxies using the very long baseline interferometry (VLBI) technique, where radio jet activities are also found in sources previously identified as RQ AGN \citep{leipski_radio_2006,white_radio-quiet_2015,herrera_ruiz_unveiling_2016,jarvis_prevalence_2019}. The importance of identifying jet activities in different quasar populations is underlined by popular galaxy evolution theories as well as recent observational evidence, where the radio jet, acting through feedback processes, dominates over other AGN energetic input (e.g., winds) in regulating the growth of galaxies and shaping the local environment in which they reside \citep{fabian_observational_2012,heckman_coevolution_2014,heckman_global_2023,kondapally_radio-agn_2024}. \par

To fully understand the cosmic evolution of radio quasar populations and feedback processes, it is important to answer key questions such as: what causes the AGN jet power to vary across such a wide scale \citep[$>3$ orders of magnitudes, e.g.][]{sikora_radio_2007}, despite little differences in their optical properties? This is especially difficult from the theoretical side since the physics of how these jets are launched and powered is still poorly understood \citep{blandford_relativistic_2019,hardcastle_radio_2020,davis_magnetohydrodynamics_2020}. Meanwhile, statistical studies of quasar properties have helped shed some light on the possible factors that may contribute to the jet production and powering. It is now believed that the jet powering mechanism is likely to be associated with a combination of accretion rate, BH mass, BH spin, and environment \citep[see the review by][and the references therein]{hardcastle_radio_2020}. \par

In this paper, we focus on the environmental factor, i.e. the role of the halo-scale environment in the triggering and powering of radio jets. Previous works have already found extensive observational evidence linking the local environmental richness to the presence of jets in radio AGN, a population that includes both high-excitation radio galaxies (HERGs) and low-excitation radio galaxies (LERGs). Radio AGN selected in previous radio surveys prefer to reside in more massive galaxies \citep[e.g.][]{best_host_2005}, and observations in the local Universe were able to pinpoint them to denser environments \citep[][]{ramos_almeida_environments_2013,ineson_link_2015,tadhunter_radio_2016}; radio AGN also show a higher incidence with clusters or galaxy groups mapped by the rich hot gas detected in the X-ray bands \citep{lin_radio_2007,dunn_radio_2010}, and are often the brightest cluster galaxies \citep[BCGs; e.g.][]{best_prevalence_2007}. Such coincidence between radio AGN and massive cosmological structures suggests that the triggering of radio jets may also be linked to their host dark matter halo. An extensive review on the topic is presented in \citet{magliocchetti_hosts_2022}.\par

However, due to limiting factors in the sensitivity and size of radio surveys, few studies to date have been able to push the environmental studies on jet production beyond the local universe ($z>1$) and into dark matter halo scales, which requires an extensive statistical sample to perform galaxy clustering analysis and estimate the host dark matter halo mass through auto-correlation or cross-correlation functions (see Section~\ref{sec:tpcf} and \ref{sec:bias} for details). A series of works \citep{lindsay_galaxy_2014,magliocchetti_clustering_2017,hale_clustering_2018} investigated the clustering properties of radio galaxies detected in 1.4 GHz through FIRST and VLA-COSMOS surveys, and consistently found that radio AGN tend to reside in haloes with $M>10^{13}h^{-1}\ M_\odot$, which is at least an order of magnitude more massive than the dark matter haloes hosting local star-forming galaxies ($M<10^{12}h^{-1}\ M_\odot$). Recent LOFAR observations have helped expand the sample size and reduce the systematics of clustering measurements of radio galaxies. \citet{hale_cosmology_2024} computed the angular clustering of the radio sources in the LOFAR Two-metre Sky Survey (LoTSS) DR2 catalogue and found a galaxy bias similar to previous results observed in other radio bands. \citet{petter_environments_2024} focused on the powerful radio sources in LoTSS DR2 with radio luminosities $L_\mathrm{150MHz}>10^{25.5}\mathrm{W}\ \mathrm{Hz}^{-1}$ out to $z\sim2$ and analysed the environments they reside in by modelling their halo occupation distribution (HOD) model. They found that luminous radio galaxies are strongly clustered at $z<2$, hosted by cluster environments with a typical halo mass of $10^{13}-10^{14}h^{-1}\ M_\odot$, and dominate the feedback processes at $z<2$. \par

Although previous clustering studies of radio AGN benefit from a (relatively) complete selection of jetted sources, the caveat lies in the lack of reliable redshift information, since a large fraction of radio sources only have faint or undetected optical counterparts \citep[e.g.][]{siewert_one-_2020,hale_cosmology_2024}. The underlying redshift distribution either needs to be constructed from photometric redshift estimations or has to rely on follow-up spectroscopy of deep-field radio observations covering a small sky area; as a result, complicated systematics lie in the clustering measurements of radio sources. \par

A good way to mitigate the difficulties of accurate redshift distributions is to study the clustering properties of radio \emph{quasars} selected with optical surveys (i.e. the high-excitation radio galaxies; HERGs). Large optical surveys with wide-field multi-object spectrographs, including the 2dF Quasar Redshift Survey \citep{croom_2df_2005} and the Sloan Digital Sky Survey \citep[SDSS;][]{myers_clustering_2007,ross_clustering_2009,shen_quasar_2009,white_clustering_2012}, have generated robust redshift measurements and completeness corrections for optically selected quasars. Therefore, quasars are well-established as robust tracers of local overdensities of the underlying dark matter distribution. The aforementioned works have reached excellent agreement where optical quasars (mostly radio-quiet radiative-mode AGN) at low to intermediate redshift ($z<3$) reside in dark matter haloes with a typical mass of $M\sim10^{12}h^{-1}\ M_\odot$ at all redshifts, which correspond to much poorer environments than the radio-selected AGN population (dominated by LERGs) mentioned above. By examining the clustering of radio-jetted optical quasars, we will know if these sources cluster like the other jetted sources (i.e. the HERGs cluster like LERGs) or like the other radiative-mode accretors (i.e. the HERGs cluster like radio quiet quasars). This will help answer the following question: is the difference in large-scale environments related to the \emph{triggering} of radio jets (i.e., all radio AGN reside in denser environments, regardless of accretion mode), or is it linked to the \emph{fueling mechanism} of radio jets (i.e., radiatively efficient AGN are in poorer environments than radiatively inefficient AGN)? Furthermore, since estimates of the black hole masses of quasars are available through their broad emission line widths, we can investigate the role of black hole mass in driving these differences. \par

To address the above questions, works including \citet{shen_quasar_2009}, \citet{mandelbaum_halo_2009}, \citet{donoso_clustering_2010}, and \citet{retana-montenegro_probing_2017} set out to compare radio-loud and radio-quiet subsamples of spectroscopically identified quasars in SDSS, where robust redshifts and completeness corrections are available. Both \citet{shen_quasar_2009} and \citet{retana-montenegro_probing_2017} used SDSS quasars with and without FIRST detections as the proxy for defining their RL/RQ quasar population, and computed the auto-correlation function of each quasar population. Their results revealed that RL quasars are always more strongly clustered than RQ quasars at all cosmic epochs, and the typical host dark matter halo masses of RL quasars are somewhat similar to those of radio galaxies ($\sim10^{13}h^{-1}\ M_\odot$). This is consistent with the halo masses measured with HOD modelling and the galaxy-galaxy lensing approach, using local ($z<0.3$) quasar samples \citep{mandelbaum_halo_2009}. Using the cross-correlation between AGN and luminous red galaxies (LRGs), within a relatively narrow redshift range ($0.4<z<0.8$), \citet{donoso_clustering_2010} reported that RL AGN (HERGs+LERGs) are more clustered than RL quasars (HERGs) at $L_\mathrm{1.4GHz}<10^{26}\mathrm{W}\ \mathrm{Hz}^{-1}$, and both populations are more clustered than the RQ LRGs. \par

All of the above studies used single-value thresholds in radio flux densities for classifying RL/RQ quasars/AGN, either the FIRST radio detection limit or cuts in radio luminosity or radio loudness. However, it is clear that a single threshold does not accurately reflect the physical nature of radio emission in different quasar populations, with the entire idea of the RL/RQ dichotomy also being challenged. Recent statistical studies with LoTSS data revealed that different components in quasar radio emission, including jet activity and host galaxy star formation (SF), evolve with quasar bolometric luminosity and redshift \citep{macfarlane_radio_2021}, which is not reflected by a single threshold. To mitigate this issue, based on the two-component model proposed in \citet{macfarlane_radio_2021},  \citet{yue_novel_2024} developed a new Bayesian model that can statistically separate the contribution of the host galaxy SF and AGN jets from the observed quasar radio emission. By fitting against the radio flux density distribution of SDSS quasars measured by LOFAR, the model recovers the underlying distribution of radio emission from host galaxy and jet, which serves as a robust proxy for defining radio quasar populations given that it is based on actual physical processes rather than single-valued cuts. Implementing the model-motivated definitions, \citet{yue_novel_2025} found that only 60\% of RL quasars classified by radio loudness defined with radio luminosities ($\mathcal{R}=L_\mathrm{144MHz}/L_{i-\textrm{band}}$) are actually dominated by jet activities, highlighting the potential systematics in the previous radio quasar clustering analysis based on the traditional bimodal classification. \par

In this work, building on the novel classification method in \citet{yue_novel_2025}, together with the excellent spectroscopic redshift coverage of the SDSS DR16 quasar catalogue and the improved sensitivity and spatial coverage of the LoTSS DR2 survey, we will investigate the clustering properties of radio quasar populations with greatly reduced systematics and unprecedented detail, and discuss their dependencies on BH mass and quasar luminosity. With our new methodology, we hope to answer the following questions: What role does the halo environment play in the powering of jets? Is there a fundamental difference in large-scale environments between AGN populations with different radio loudness? Is there a minimum halo mass required to trigger quasar radio jets? \par

This paper is structured as follows: Section~\ref{sec:data} describes the construction of the quasar sample set used in our analysis. Section~\ref{sec:method} outlines the methodology used in the clustering analysis, including a physically motivated definition of radio quasar populations, and ways to measure the spatial clustering of different quasar populations through the two-point correlation function (TPCF). Section~\ref{sec:result} presents our clustering measurements and their dependencies on BH mass and quasar luminosities. Section~\ref{sec:bias} calculates the quasar bias from our clustering measurements and provides order-of-magnitude estimations of the corresponding dark matter halo masses. Finally, Section~\ref{sec:discussion} discusses the insights into jet production and triggering mechanisms based on our measurements. \par

Throughout this work, we assume a flat $\mathrm{\Lambda CDM}$ cosmology with matter density $\Omega_m=0.3$ and the cosmological constant $\Omega_\Lambda=0.7$. We assume a Hubble constant $H_0=70\ \textrm{km}\ \textrm{s}^{-1}\ \textrm{Mpc}^{-1}$ (i.e., $h=0.7$), and $\sigma_8$, the rms mass fluctuation amplitude in spheres of size $8\ h^{-1}\ \textrm{Mpc}$, has a value of $\sigma_8=0.84$. All distances in this work are comoving distances expressed in units of $h^{-1}\ \textrm{Mpc}$.

\section{Data}
\label{sec:data}

\subsection{Extended Baryon Oscillation Spectroscopic Survey (eBOSS) quasars }
\label{sec:data_sdss}

We built our parent quasar sample from the completed SDSS-IV extended Baryon Oscillation Spectroscopic Survey (eBOSS) quasar catalogue \citep{ross_completed_2020}; its well-defined random catalogue can accurately map the detection systematics, which is critical to robustly determining the variance in quasar clustering measurements.  The eBOSS programme began in July 2014 and was carried out with double-armed spectrographs mounted on the Sloan Foundation Telescope at Apache Point Observatory \citep{gunn_25_2006}. The quasars in the eBOSS programme are selected based on the criteria presented in \citet{myers_sdss-iv_2015}: for quasars with redshift $0.8<z<2.2$, the selection used optical photometric data from SDSS-I/II/III, together with a mid-infrared cut based on Wide-field Infrared Survey Explorer \citep[WISE;][]{wright_wide-field_2010} observations; quasars with redshift $z>2.1$ are complemented with an additional selection based on \ion{Ly}{$\alpha$} forest measurements. The final eBOSS quasar sample covers $4699\ \textrm{deg}^2$ of the sky area, with 343,708 quasars identified in the redshift range $0.8<z<2.2$ and 72,667 quasars with redshifts $2.2<z<3.5$. The redshifts of the eBOSS quasars are determined from observed SDSS spectra, using the algorithm described in \citet{lyke_sloan_2020}. Although the entire redshift range of the eBOSS quasars spans from $z=0.8-3.5$ (see Figure~\ref{fig:opticalproperty}), we follow the recommendations of \citet{ross_completed_2020} and \citet{hou_completed_2021} and concentrate on the eBOSS quasar sample that spans the redshift range of $0.8<z<2.2$, since quasars in this redshift range are uniformly selected and, therefore, optimised for large-scale structure studies. \par

Prior to our analysis, we must correct for any systematics that affect the completeness of quasar detections in our target field. We adopt the weights described in \citet{rezaie_primordial_2021} for the completeness corrections in this paper. The weights in \citet{rezaie_primordial_2021} follow the definition in \citet{ross_completed_2020}, which characterises four different types of systematics:

\begin{enumerate}
\item $w_\textrm{sys}$ is introduced to mitigate the imaging systematics, including trends in the $g$-band depth, Galactic extinction, sky background and seeing;
\item $w_\textrm{cp}$ is introduced to correct for fibre collision. In each observation, the minimum angular distance between two quasar targets is limited by the physical structure that supports the fibres, which has a projected size of $62\arcsec$. Quasar targets that fall within the angular separation limit are corrected with the close pair (fibre collision) weight $w_\textrm{cp}$.
\item $w_\textrm{noz}$ is introduced to correct for the efficiency of the redshift measurement, which varies between different fibres, showing a dependence on the fibre ID number. Fibres close to the edge of the spectrograph tend to have a lower redshift efficiency.
\item $w_\textsc{fkp}$ is applied to minimise the cosmic variance of source detection, where $w_\textsc{fkp}=(1+P_0n(z))^{-1}$ \citep{feldman_power-spectrum_1994}, with $P_0=6000h^{-3}\textrm{Mpc}^3$ being the value of the eBOSS quasar power spectrum $P(k)$ at $k=0.14h\textrm{Mpc}^{-1}$, and $n(z)$ being the expected comoving volume number density calculated from the selection function in each redshift bin \citep{neveux_completed_2020}.
\end{enumerate}

The final weight applied to each object is, therefore, defined as $w_\textrm{tot}=w_\textsc{fkp}w_\textrm{sys}w_\textrm{cp}w_\textrm{noz}$. \citet{rezaie_primordial_2021} improved the weight estimations in \citet{ross_completed_2020} by introducing a neural network-based method that mitigates fluctuations in the density field caused by spatial variations in the quality of the imaging data used in the target selection. Such non-linear imaging systematics will cause a steepening in the two-point correlation function (see Section~\ref{sec:tpcf}) at the largest distance scales. As a result, adopting the weight presented in \citet{rezaie_primordial_2021} allows robust measurement of larger-scale quasar clustering ($k<0.01\ h\ \textrm{Mpc}^{-1}$). This is especially relevant to our analysis, since the clustering scales of jet-dominated quasars are often much larger than those of the entire quasar population \citep[e.g.][]{shen_quasar_2009,retana-montenegro_probing_2017}, where updated weights become necessary to make a robust measurement. \par

In addition to the sky coordinates and redshift measurements provided in \citet{ross_completed_2020}, we have also compiled the physical properties for eBOSS quasars, including \emph{i}-band luminosities ($\mathcal{M}_i$) and black hole masses ($M_\textrm{BH}$). These values are adopted from the SDSS DR16Q spectroscopic catalogue \citep[][]{wu_catalog_2022}. To obtain the BH mass, \citet{wu_catalog_2022} fit the SDSS quasar spectra with a global continuum + emission line model, using the public PyQSOFIT code \citep[][]{shen_sloan_2019}{}{}. BH masses were measured from single-epoch emission line FWHMs of \ion{Mg}{ii} (for $0.8<z<2.0$) and \ion{C}{iv} (for $2.0<z<2.2$), using the virial relation provided in \citet{schneider_sloan_2010}. \citet{wu_catalog_2022} reported a mean discrepancy of $0.1-0.2$ dex between BH masses measured with different emission line estimators on the same object, with larger offsets seen in low-luminosity quasars where the \ion{C}{iv} profile exhibits a narrow core; however, since the BH masses are estimated with only one type of emission line in each of our redshift bins, we still include the full redshift range when studying the BH mass dependence (Section~\ref{sec:result_mass}). All aforementioned physical properties will be used to distinguish radio quasar sub-populations in our clustering analysis.

\subsection{Matching LoTSS data with eBOSS quasars}

\subsubsection{LOFAR Two-metre Sky Survey (LoTSS)}
\label{sec:data_lotss}

\begin{figure}     
    \includegraphics[width=\columnwidth]{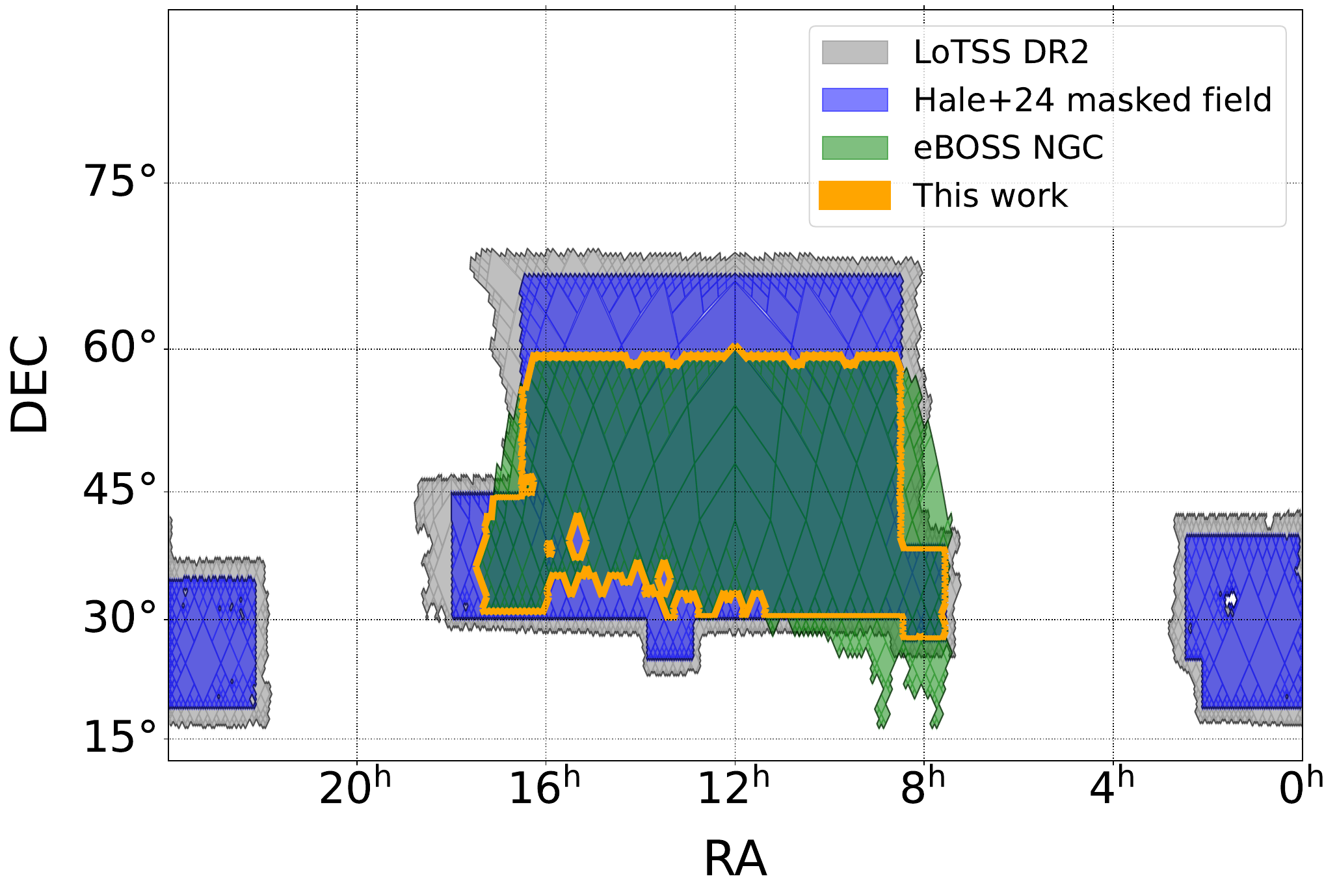}
    \caption{A comparison of the sky coverages from LoTSS DR2 (grey) and eBOSS (green) survey catalogues. The masked LoTSS DR2 field for reduced flux variation systematics is shown in blue, according to the definition in \citet{hale_cosmology_2024}. The orange line encompasses the sky area studied in this paper, which is the overlap between the masked LoTSS DR2 area and the eBOSS survey area in the north galactic cap. Note that the holes in the sky area are due to the uneven spatial coverage in the eBOSS survey.}
    \label{fig:lotssdr2sky}
\end{figure}

We obtain the radio flux density measurements for the eBOSS quasars from the LOFAR Two-metre Sky Survey \citep[LoTSS;][]{shimwell_lofar_2017, shimwell_lofar_2019, shimwell_lofar_2022}, a wide-field low-frequency radio imaging survey conducted with the LOFAR HBA (high-band antenna). The goal of LoTSS is to survey the entire northern sky in the 120-168 MHz radio band, with a designed rms noise level of $\lesssim100\mu\mathrm{Jy\ beam}^{-1}$, an angular resolution of $6 \arcsec$, and a positional accuracy within $0.2 \arcsec$. In this study, we use the LoTSS DR2 catalogue \citep{shimwell_lofar_2022}, as it provides a near-complete coverage of the eBOSS field. LoTSS DR2 covers a sky area of 5,720 $\mathrm{deg}^2$ and achieves a median rms noise of $83\mu\mathrm{Jy\ beam}^{-1}$. The current LoTSS DR2 catalogue comprises approximately 4,400,000 radio-detected sources. The sky coverage of LoTSS DR2 (grey), in comparison with the eBOSS field (green), is shown in Fig.~\ref{fig:lotssdr2sky}. \par

In LoTSS DR2, the radio sources were identified from the LoTSS images using the Python Blob Detector and Source Finder \citep[PyBDSF;][]{mohan_pybdsf_2015}, based on the peak radio flux densities surpassing the $5\sigma$ threshold in the LoTSS DR2 images. Although PyBDSF is effective at detecting regions of radio emission, it does not always properly associate them with physical sources. To mitigate such mis-associations, a combination of statistical methods and visual examination (thanks to the LOFAR Galaxy Zoo) was applied to radio sources with flux densities $>4\ \mathrm{mJy}$ to ensure that the radio catalogue reflects the actual distribution of radio sources \citep[][]{hardcastle_lofar_2023}. The catalogued LoTSS DR2 sources were also cross-referenced with their optical/infrared counterparts in the WISE \citep{wright_wide-field_2010} and DESI Legacy Imaging \citep{dey_overview_2019} surveys, using a similar methodology described in \citet{williams_lofar_2019} and \citet{kondapally_lofar_2021}. \par

\subsubsection{Radio Flux Densities from LoTSS}
The radio flux densities of the eBOSS quasars are extracted from both the LoTSS DR2 catalogue and the original LoTSS mosaic. Here we briefly summarise our approach, while the details of the process are described in \citet{yue_novel_2024}: \par

\begin{enumerate}
    \item Firstly, we cross-matched the positions of quasars in the eBOSS catalogue with the positions of radio sources in the source-associated LoTSS DR2 catalogue (which adopts a $5\sigma$ detection limit) using a matching radius of $1.5\arcsec$. These sources are automatically identified with PyBDSF, and their radio flux densities are adopted from the values recorded in the LoTSS DR2 catalogue, measured using the methods described previously. Note that whenever available, we used the coordinates from the cross-matched optical catalogues of \citet{hardcastle_lofar_2023} for the positions of the radio counterparts. \par
    
    \item For sources not detected in the LoTSS DR2 catalogue, we performed forced photometry on the LoTSS mosaics using the method described in \citet{gloudemans_low_2021}. This approach is motivated by the assumption that most quasars are unresolved in LoTSS DR2 images, and the astrometric uncertainty in LoTSS DR2 ($\lesssim0.2\arcsec$) is significantly smaller than the LoTSS pixel scale ($1.5\arcsec$). The extracted radio flux density is considered to be the highest pixel value within a $3\mathrm{px}\ \times\ 3\mathrm{px}$ aperture centred on the quasar's optical position in the SDSS catalogue. This aperture size accounts for the possibility that the source's radio emission is misaligned with the location of its optical emission. The uncertainty in the flux density is calculated based on the standard deviation of pixel values in a $100\mathrm{px}\ \times\ 100\mathrm{px}$ cut-out region surrounding the central pixel. As discussed in \citet{yue_novel_2024}, although these individual measurements are rather noisy, they retain valuable information about the distribution of radio flux densities of the population as a whole.
\par
\end{enumerate}

\subsection{Building a LoTSS-eBOSS quasar sample}
\label{sec:data_build}

We built our quasar sample set based on the extracted LoTSS radio flux density measurements, together with the redshift and RA/Dec information of the parent eBOSS quasar samples. Furthermore, we have imposed positional constraints on the data as recommended by \citet{hale_cosmology_2024}, removing fields where the LOFAR pointings were not mosaiced together. As discussed in \citet{shimwell_lofar_2022}, the flux scale of measurements can vary across an individual LOFAR pointing, due to differences in the model of the primary beam across the field of view. Such flux scale variation will introduce systematics to the completeness of radio detections (and non-detections), but can be reduced where LOFAR pointings have been mosaiced together. To mitigate this systematics, we made the same positional cuts to the data and random samples as described in \citet{hale_cosmology_2024}, removing the outer edge of the LoTSS DR2 field. The final LoTSS field used in this work is marked in blue, as shown in Figure~\ref{fig:lotssdr2sky}. \par

\begin{figure}     
    \includegraphics[width=\columnwidth]{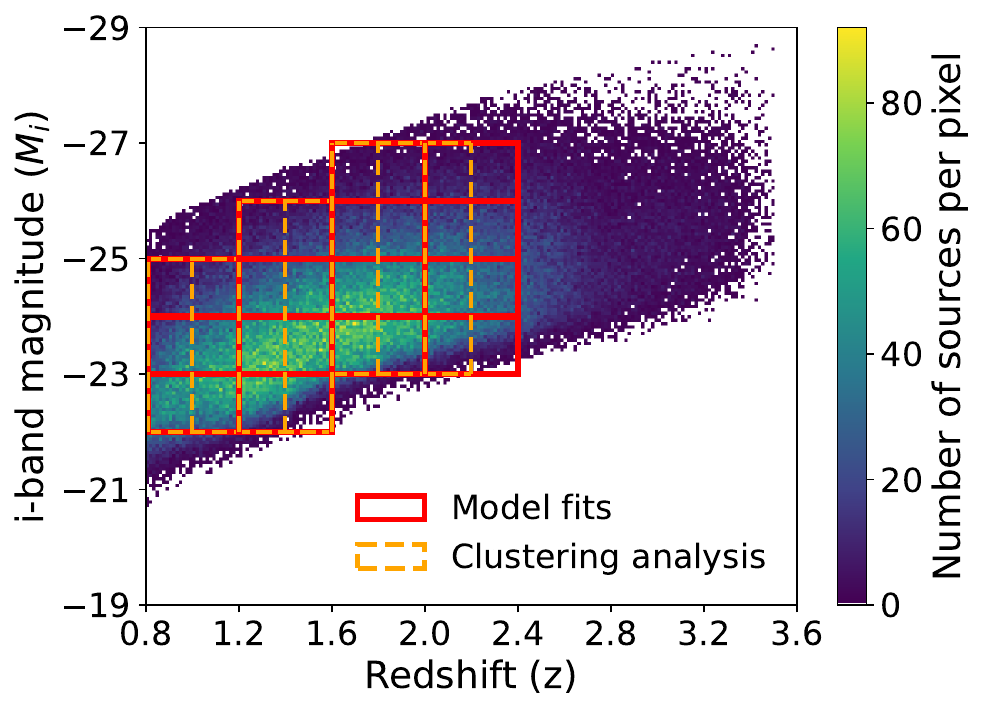}
    \caption{Distribution of eBOSS quasars in the $\mathcal{M}_i-z$ plane. The red solid grids show the $\mathcal{M}_i-z$ grids with at least 3,000 quasars within the redshift range $0.8<z<2.2$. We use these grids to obtain best-fits of the two-component model, as well as defining the radio quasar populations in this work. The dashed orange grid outlines the binning used for clustering analysis, with smaller redshift binning and an upper limit of $z<2.2$. We combine the quasars along the $\mathcal{M}_i$ axis and compute the clustering signals in each redshift slice.}
    \label{fig:opticalproperty}
\end{figure}

The quasar sub-samples are characterised by their distribution in the $\mathcal{M}_i-z$ plane, as shown in Fig.~\ref{fig:opticalproperty}. The best-fits of our two-component model (see the discussion in Section~\ref{sec:model} for details) are determined in each of the $\mathcal{M}_i-z$ grids with at least 3,000 sources (outlined in red solid lines) and are then used to classify quasars based on their radio emission. This is consistent with the binning used in \citet{yue_novel_2025}, in order to gather enough source counts for statistical analysis. Note that the minimum number of sources adopted here is greater than the recommended lower limit of 1,000 \citep{yue_novel_2024}; the increase in the requirement of source count is to maximise the robustness of our model-based classification. For quasar clustering measurements, we rebin our samples into smaller $\mathcal{M}_i-z$ grids, as shown in orange dashed lines. This is because clustering measurements require fewer source counts, and therefore we can have higher resolution in the redshift space. We use the two-component model to define quasar subsamples in each grid cell (see Section~\ref{sec:model}); then we combine the grid cells along the $\mathcal{M}_i$ axis and measure the clustering signal within each redshift slice. The final sample includes 225,382 quasars, of which 189,979 quasars are used in the analysis in this work. \par

\section{Method}
\label{sec:method}

\subsection{Classifying radio quasar population by physical processes}
\label{sec:model}

In contrast to previous works where quasars are split into two distinct populations -- radio-loud (RL) quasars and radio-quiet (RQ) quasars -- using a fixed threshold, before measuring and comparing their clustering properties, this work uses the two-component Bayesian model proposed in \citet{yue_novel_2024} to separate and quantify radio emission from the host galaxy and AGN activity. Instead of a binary classification based on radio loudness or radio (non-)detection, we bin the quasars on the basis of their physical process, characterised by the jet fraction fitted through the two-component model. This is only possible due to the unique depth and spatial coverage of the LoTSS survey, which provides enough statistics in the measured radio flux densities for such modelling. \par

Our two-component model takes in the distribution of observed quasar radio flux densities within a reasonably small grid cell within the parameter space defined with bolometric luminosity (here traced by the SDSS \emph{i}-band magnitude $\mathcal{M}_i$) and redshift ($z$), while assuming that \emph{every} quasar hosts radio emission coming from the host galaxy SF activity and AGN activities. Following the paradigm in \citet{macfarlane_radio_2021}, we assumed a log-Gaussian distribution for the underlying probability distribution function (PDF) of the host galaxy SF component in quasar radio emission ($P_\mathrm{SF}$), centred on the average radio SFR within the grid cell, while assuming a single power-law distribution for the PDF of the AGN component ($P_\mathrm{AGN}$; mostly tracing jet activities at a wide range of powers), extrapolated from the radio-loud population. \par

\begin{figure*}     
    \includegraphics[width=0.6\linewidth]{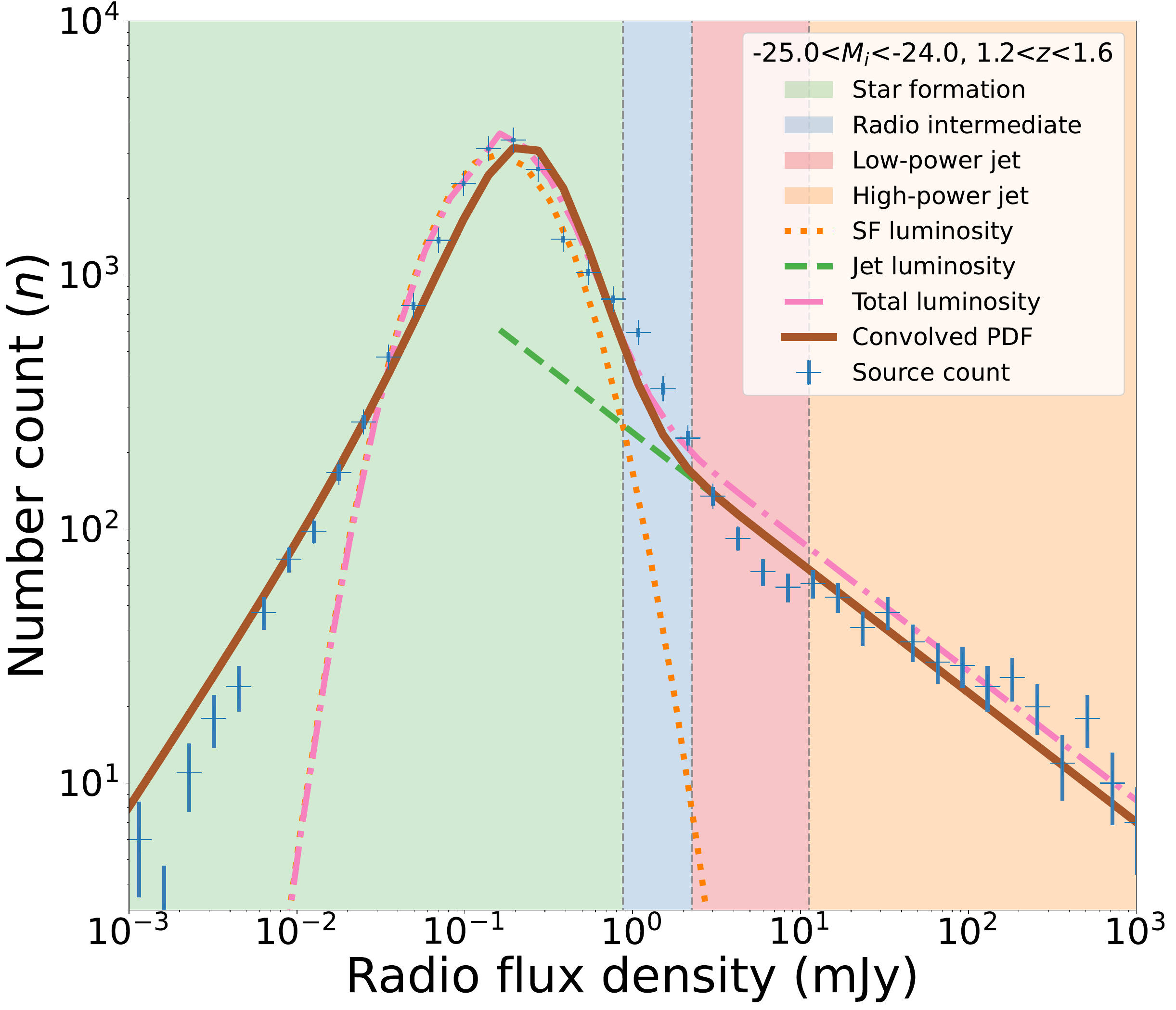}
    \caption{Radio flux density distribution and the corresponding model best-fit in one of the representative $\mathcal{M}_i-z$ bins explored in this study. The orange dotted line and green dashed line represent the SF component from the host galaxy activity (log Gaussian) and the jet component from the AGN activity (single power-law) respectively. The pink dashed-dotted line shows the combined PDF of the two-component radio flux density distribution model, and the brown solid line shows the probability density function (PDF) after convolving with a Gaussian uncertainty characterising noise in the flux density measurements. This convolution spreads the original PDF at the faint end out to negative flux density values, leading to an offset between the convolved and original PDFs at flux densities lower than $\sim100\mathrm{\mu Jy}$. The distribution of measured radio flux densities is shown as blue crosses, indicating a good match between the actual distribution and the best-fit model result. The coloured areas highlight the different radio quasar populations defined with their dominant sources of radio emission, based on the best-fit of our model. From left to right are populations dominated by: host galaxy SF (green), mixture of SF and AGN activities (blue; radio intermediate region with no dominant source), low-power AGN jets (red), and high-power AGN jets (orange). }
    \label{fig:pdf}
\end{figure*}

Figure~\ref{fig:pdf} shows the PDFs of different components calculated with the best-fit model parameters, within one of the $\mathcal{M}_i-z$ grids that we explored in this work. The orange dotted line shows the fitted underlying radio flux density distribution of host galaxy SF activity ($P_\mathrm{SF}$), while the green dashed line shows the underlying distribution of AGN activity ($P_\mathrm{AGN}$). Combining these two components, we have the PDF for the total \emph{underlying} radio flux density distribution for the quasar population (pink dash-dotted line). This is then convolved with the observational noise to produce a model PDF of the \emph{observed} quasar radio flux density distribution, which is shown as the solid brown line. We then fit our final convolved PDF to the data with a Bayesian approach, based on the methodology developed in \citet{roseboom_cosmic_2014} and \citet{yue_novel_2024}, to get the set of best-fit model parameters that allow us to reconstruct the distributions of radio emission from SF and AGN activities from the observed data. Our convolved PDF shows good agreement with the observed quasar radio flux densities (blue crosses) across all $\mathcal{M}_i-z$ grids, indicating that our model provides a good description of the quasar sample. \par

Having the ability to decompose the underlying sources of quasar radio emission, our two-component model opens up the possibility of an alternative definition of radio quasar populations, which is based on actual physical processes and takes into account variations across parameter space, rather than binary radio-loudness cuts. In this work, within a given $\mathcal{M}_i-z$ grid, we split the quasar samples into four different populations based on the dominant sources of their radio emissions, characterised by the underlying PDFs of SF and AGN components, as shown in Figure~\ref{fig:pdf}. 

The four radio quasar populations are defined as follows:

\begin{enumerate}
    \item SF-dominated quasars (SFQs). These quasars occupy the leftmost region of the radio flux density distribution in Figure~\ref{fig:pdf} and are highlighted in green. In this parameter space, the underlying PDF of the SF component is higher than that of the AGN component, indicating that radio emission from host galaxy star formation is the dominant radio continuum emission mechanism. The upper flux density limit of this quasar population is defined as $S_\textrm{eq}$ where $\mathcal{P}_\textrm{SF}(S_\textrm{eq})=\mathcal{P}_\textrm{AGN}(S_\textrm{eq})$. This population consists of 167,009 quasars ($87.9\%$ of the entire population).
    \item Radio-intermediate quasars (RIQs). These quasars have flux densities higher than those of SF-dominated quasars, and are highlighted in blue in Figure~\ref{fig:pdf}. In this parameter space, there is no clear dominance of either the SF or the AGN component, since $\mathcal{P}_\textrm{SF}$ is smaller than $\mathcal{P}_\textrm{AGN}$, yet the convolved PDF still deviates from the single power law $\mathcal{P}_\textrm{AGN}$. The upper limit of this quasar population is defined as $S_\textrm{lo-jet}$ where $\mathcal{P}_\textrm{SF}(S_\textrm{lo-jet})
/\mathcal{P}_\textrm{AGN}(S_\textrm{lo-jet})=0.05$. This population consists of 14,903 quasars (7.8\%).
\item Low-power jet-dominated (lo-jet) quasars. These quasars reside in the red area in Figure~\ref{fig:pdf}, where their radio flux densities are dominated by the single power law jet component ($\mathcal{P}_\textrm{SF}<0.05\mathcal{P}_\textrm{jet}$), but their radio powers are not as strong as high-power jet-dominated quasars. Depending on the $\mathcal{M}_i-z$ grid they reside in, these quasars have typical radio flux densities of $S_\textrm{144MHz}=1-10\ \textrm{mJy}$, which lies at the border of the detection limit of the FIRST survey ($S_\textrm{1.4GHz}\approx1\ \textrm{mJy}$\footnote{This corresponds to $S_\textrm{144MHz}\approx1.98\ \textrm{mJy}$, assuming jet dominated ($S_\nu\propto\nu^{-0.3}$), or $S_\textrm{144MHz}\approx4.91\ \textrm{mJy}$, assuming SF dominated ($S_\nu\propto\nu^{-0.7}$).}). However, these quasars are not always radio-loud according to classical definitions (e.g. $R\equiv S_\textrm{5GHz}/f_\textrm{4400\AA}>10$; see the discussion below). The upper flux density limit of this population is defined as $S_\textrm{hi-jet}$ where $\mathcal{P}_\textrm{SF}(S_\textrm{hi-jet})
/\mathcal{P}_\textrm{AGN}(S_\textrm{hi-jet})=10^{-5}$. This population consists of 4,897 quasars (2.6\%).
\item High-power jet-dominated (hi-jet) quasars. These quasars occupy the orange area in Figure~\ref{fig:pdf}, and are the most powerful radio sources in our sample, with radio flux densities $S>S_\textrm{hi-jet}$. They are a pure sample of what are traditionally defined as RL quasars, with $R\gtrsim30$. This population consists of 3,170 quasars ($1.7\%$).
\end{enumerate}

The motivation behind defining quasar populations based on the two-component model is that the binary split of RL versus RQ quasars fails to address the underlying production mechanism of quasar radio emission, regardless of the radio loudness threshold used. The `division point' between RL and RQ also varies with $\mathcal{M}_i$ and $z$, which is not captured by the traditional binary separations \citep[][]{yue_novel_2025}. 

\begin{figure}
  \centering
  \includegraphics[width=\linewidth]{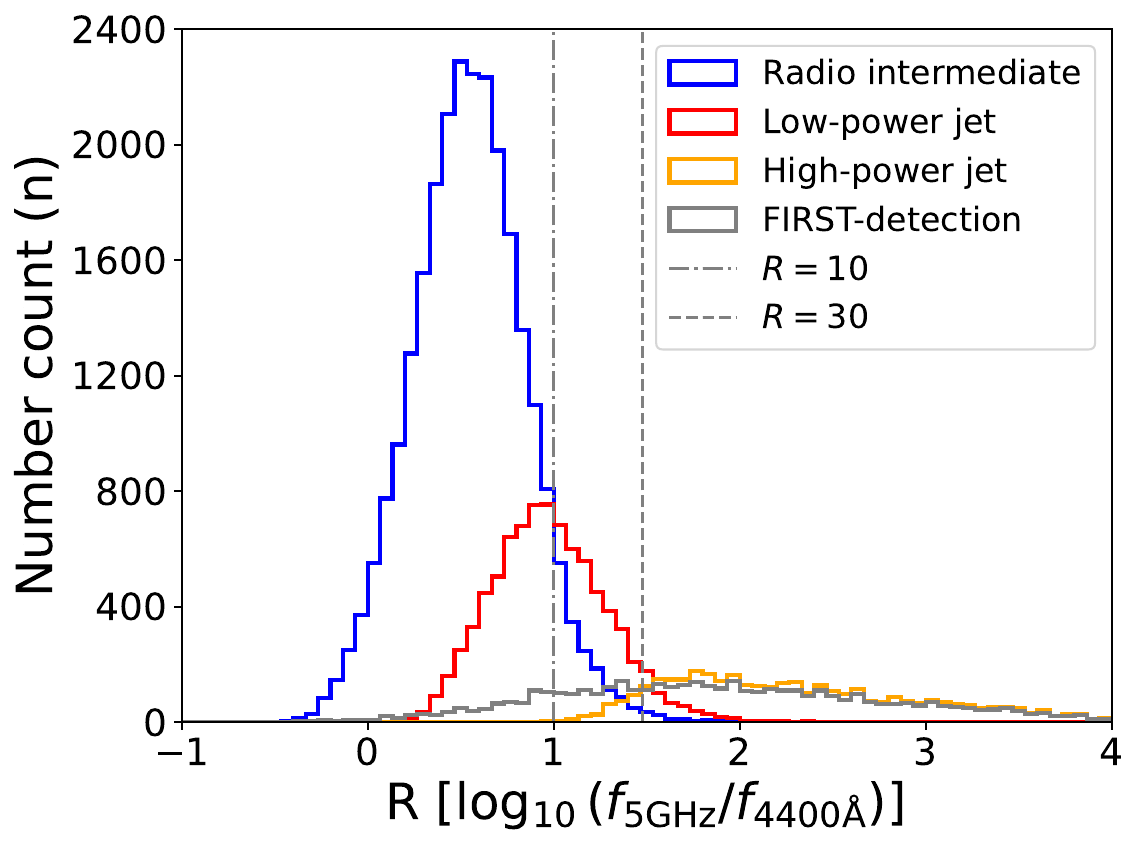}
  \caption{The distribution of radio loudness, defined with $R=\log_{10}(S_\mathrm{5GHz}/f_\mathrm{4400\AA})$, across different quasar populations defined in this study. The vertical dashed-dotted line and dashed line marks two different thresholds used to define radio-loud quasars in previous studies: $R=10$ and $R=30$. The grey histograms marks the distribution of FIRST-detected quasars, another definition of the radio-loud quasar population. The optical $f_\textrm{4400\AA}$ flux density is calculated from the luminosity at 3000\AA ($L_\textrm{3000\AA}$) in the SDSS DR16Q catalogue, assuming a spectral slope of $f_\lambda\propto\lambda^{0.5}$. The radio $S_\mathrm{5GHz}$ flux density is calculated from LoTSS 144 MHz or FIRST 1.4 GHz (for sources with FIRST-detection) flux densities assuming a spectral slope of $S_\nu\propto\nu^{-0.3}$.}
  \label{fig:rl_dist}
\end{figure}

To demonstrate this, Figure~\ref{fig:rl_dist} compares the radio loudness values defined with $R\equiv S_\textrm{5GHz}/f_\textrm{4400\AA}$ \citep[see e.g.][]{kellermann_vla_1989,jiang_radio-loud_2007} in the RIQ, lo-jet, and hi-jet quasar populations as defined above. Here, since these quasars are assumed to have a moderate-to-strong jet component, the $f_\textrm{5GHz}$ radio flux density is calculated from the 144 MHz LOFAR radio flux density using the typical spectral index for \emph{radio-loud quasars}, $S_\nu\propto\nu^{-0.3}$ \citep{gurkan_lotsshetdex_2019}. Meanwhile, $f_\textrm{4400\AA}$ is calculated from the luminosity at 3000\AA ($L_\textrm{3000\AA}$) given by the SDSS DR16Q catalogue \citep{wu_catalog_2022}, assuming a spectral slope of $f_\lambda\propto\lambda^{0.5}$. The dashed-dotted and dashed vertical lines mark the two criteria used in \citet{jiang_radio-loud_2007} to select the RL quasars: $R>10$ and $R>30$. Most jet-dominated quasars are included in the RL population under the definition of $R>10$, yet a significant portion of lo-jet quasars are still left out by this selection, while including a reasonable amount of contaminant RI quasars. On the other hand, the hi-jet population is selected by $R>30$ with almost no contamination from the radio-intermediate population; therefore, the $R>30$ criterion acts as a benchmark for the traditionally defined RL quasars. \par

Other radio quasar clustering studies including \citet{shen_quasar_2009} and \citet{retana-montenegro_probing_2017} used FIRST detection ($S_\textrm{1.4GHz}>1\ \textrm{mJy}$) to define the RL quasar population. This population is shown by grey histograms in Figure~\ref{fig:rl_dist}, whose radio loudness is calculated from 1.4 GHz radio flux densities presented in the FIRST catalogue (FIRST catalogue reference) instead of 144 MHz flux densities, assuming the same spectral slopes; it has an average radio loudness of $R\sim100$, and includes all hi-jet quasars. However, the FIRST-detected quasars are still a mixed population of different jet fractions, spanning from RI to hi-jet quasars (note that RI quasars do include a subset of traditionally-defined radio-quiet population). The hi-jet quasars only make up about half of the population, while a significant portion of $R\sim10$ objects (or lo-jet quasars) are still undetected in FIRST \citep[see also][]{shen_quasar_2009}. \citet{petter_environments_2024} measured the clustering properties of LOFAR-detected radio galaxies and used $L_\textrm{150MHz}\gtrsim10^{25.25}\ \textrm{W}\ \textrm{Hz}^{-1}$ to define their RL quasar sample; the typical radio loudness for this population is $R\gtrsim10$, which still hosts a mixture of different radio emission mechanisms due to the contaminant RI quasars. \par

As LOFAR probes into much fainter radio luminosities, the two-component-model classification is so far the most accurate reflection of the dominant source of radio emissions in different quasar populations. Therefore, in the following sections, we focus on the clustering properties of these four quasar populations. \par

\subsection{Selection bias in quasar sub-populations}
\label{sec:selection_bias}

While the systematics in our quasar parent sample are well modelled by the random sample and weights (see Section~\ref{sec:data_sdss}) from the eBOSS catalogue, additional selection bias may arise in classifying quasars into sub-populations based on their radio emission. Since we performed forced aperture extractions (see Section~\ref{sec:data_lotss}) to obtain radio flux density measurements for quasars with radio emission below the $5\sigma$ detection limit of LoTSS, for the faintest objects the fractional uncertainties in their radio flux density measurements can be quite large. This could potentially move quasars between different sub-populations within a $\mathcal{M}_i-z$ grid cell. \par

The limiting factor in such systematics from quasar classification therefore depends on how the noise level in the radio detections compares to the radio flux density limits between different quasar sub-populations. If all quasars within the sub-population have radio flux densities above the completeness limit of LoTSS detections (i.e., can be robustly identified by PyBDSF and are free from field-dependent imaging systematics), then the measurement uncertainty will have minimal effect on the completeness of this quasar sub-population, since the sub-sample is defined with a radio selection that is complete at this flux density level. Note that our sample will be more complete at the original completeness limit of the LoTSS survey since we have also performed forced photometry for objects not picked up by PyBDSF (although the measured flux densities may not be as reliable for extended sources). \par

To be conservative, we adopt the completeness limit provided in \citet{hale_cosmology_2024}, where they examined the completeness of the PyBDSF detected LoTSS sources in the DR2 field using image plane completeness simulations in \citet{shimwell_lofar_2022}. The completeness of LoTSS detections is strongly dependent on S/N with a small scatter, while also showing a weaker correlation with the observed flux density. The systematics in S/N are caused by PyBDSF using the same S/N threshold to detect sources across different pointings, while the rms noise level varies from field to field; therefore the completeness level will be lower in pointings with a higher rms noise. The systematics in flux density can be tied to the possible correlation between intrinsic size distribution of the sources and their flux densities, since for larger sources, the neighbouring pixels can also help in pushing the peak pixel above the detection limit. This leads to AGN activities, which are more likely to host extended emission from jet activities and are often brighter than SF activities, being easier to detect by PyBDSF. Having already removed the edge fields where the flux scale variations are systematically larger, we adopt the conservative cut of $\mathrm{S/N>7.5}$ ($7.5\sigma$ detection limit) as proposed in \citet{hale_cosmology_2024}. 

We examined the radio flux densities with which we make the cuts for each quasar population within each $\mathcal{M}_i-z$ grid. Except for grid cells with redshift and \emph{i}-band magnitude ranges of $[1.2<z<1.6,-23<\mathcal{M}_i<-22]$, and $[2.0<z<2.4, -24<\mathcal{M}_i<-23]$, the radio flux density lower limits of the RI quasars ($s_\textrm{eq}$) are all above the $7.5\sigma$ detection limit of LoTSS-wide, considering a median rms noise level of $\sigma_s=74\ \mu\mathrm{Jy}\ \mathrm{beam}^{-1}$ across the field of view \citep[][]{shimwell_lofar_2022}, while the $S_\mathrm{eq}$ for the two aforementioned grid cells are $0.46\ \mathrm{mJy}$ and $0.41\ \mathrm{mJy}$, respectively. We therefore removed these two grid cells from our analysis and consider the selection of the RI quasar population to be complete using the S/N criteria.\footnote{We have examined the choice of rms noise level in our S/N criteria by removing the fields with rms noise level above $\sigma_s=100\ \mu\mathrm{Jy}\ \mathrm{beam}^{-1}$, before performing the same analysis included in this work. With this updated cut in rms noise, all RI quasars are \emph{securely} above the $5\sigma$ detection threshold. There is no significant change to any of the results; therefore, we stick to the current data processing approach to maximise the number statistics.} As discussed above, the RIQ population is less affected by the systematics related to flux density since they are unlikely to host extended jets, and the lower flux density limits of the lo-jet quasars across the grid cells are all close to ($>1.0\ \mathrm{mJy}$) or above $1.5\ \mathrm{mJy}$. Therefore, we do not impose further corrections on the completeness of our jet-dominated quasar populations. Combining the additional S/N and flux density cuts to the classification criteria, we ensure a robust division between RI quasars, lo-jet quasars, and hi-jet quasars through our definition. \par

\subsection{Two-point correlation function}
\label{sec:tpcf}

The two-point correlation function (TPCF; $\xi(r)$) describes the excess probability of finding a quasar pair with a separation $r$, with respect to a homogeneous distribution. In practice, this is calculated by pairing the spatial distribution of the actual quasar data set with a generated random catalogue containing a set of points with the same selection function, angular geometry, and redshift distribution compared to the real dataset. In this work, we use the random catalogue that accompanies the eBOSS QSO catalogue, including the spatial coordinates, redshifts, and weights assigned to the random sample. The detailed process of generating the random catalogue for eBOSS quasars is described in \citet{ross_completed_2020}, with the updated weights presented in \citet{rezaie_primordial_2021} following the same method as described in Section~\ref{sec:data_sdss}. We applied the same positional cuts to the random catalogue as described in Section~\ref{sec:data_build}. We emphasise that after applying the additional criteria in the quasar classification proposed in the previous section, the radio selection effect no longer needs to be taken into account when constructing random samples for different quasar sub-populations, since the imaging systematics of radio observations does not affect the classification of sources. We therefore apply a uniform random for every quasar sub-population. \par

To compute the TPCF, we use the Landy-Szalay minimum variance estimator \citep{landy_bias_1993}:

\begin{equation}
    \xi(r)=\frac{DD(r)-2DR(r)+RR(r)}{RR(r)},
\end{equation}

where $DD$, $DR$, and $RR$ are the number of quasar pairs with distance $r$ selected within the data catalogue, between the data and the random catalogue, and within the random catalogue, respectively. The pair counts are then normalised by the number densities of actual and random quasars. \par

Following traditional methods \citep[e.g.][]{croom_2df_2005,ross_clustering_2009,white_clustering_2012,eftekharzadeh_clustering_2015}, we used a single power-law function to model the TPCF: $\xi_s(s)=(s/s_0)^{-\gamma}$, where $s_0$ is defined as the correlation length in the redshift space, since we directly calculate the comoving line-of-sight distances from eBOSS redshifts, following conventional notation. We then applied a chi-square method to obtain the best-fit TPCF parameters. We use the software package \textsc{TreeCorr} \citep{jarvis_skewness_2004,jarvis_treecorr_2015} to measure TPCF and its uncertainties in the 3D comoving space. The separations of quasar pairs are comoving distances calculated from RA, DEC, and comoving line-of-sight distances as inputs. TPCFs are measured in logarithmic comoving distance bins, with a bin width of $\Delta s/(h^{-1}\ \textrm{Mpc})=0.08\ \textrm{dex}$. \par

Since the TPCF signal has a real variation across the field which will dominate the total variance at large scales, one needs to split the field into patches and calculate the variation of measurements across different patches to estimate the overall uncertainty of TPCF best-fits. Following the analysis in \citet{hale_cosmology_2024}, we choose the jackknife resampling method \citep[e.g.][]{norberg_statistical_2009} to derive the uncertainties of the TPCF. In this approach, the entire sky coverage of our dataset is divided into 150 patches, and different TPCFs are calculated when each patch is removed one at a time; the uncertainty is therefore determined by the variance of these TPCF measurements. \citet{hale_cosmology_2024} demonstrated that this number of jackknife patches produces representative error estimations with the LoTSS DR2 sky coverage. In this work, we estimate the TPCF variance with the diagonal components of the covariance matrix. We acknowledge that not using the full covariance matrix would fail to account for the intrinsic correlation between the TPCF values measured at different correlation lengths. Some studies, e.g. \citet{hale_cosmology_2024}, report lower measured bias values when using the full covariance matrix as compared to only using the diagonal elements; however, they argued that the values are typically consistent to within $1-2\ \sigma$. \par

\section{Clustering Measurements}
\label{sec:result}

\begin{figure}
    \centering
    \includegraphics[width=1\linewidth]{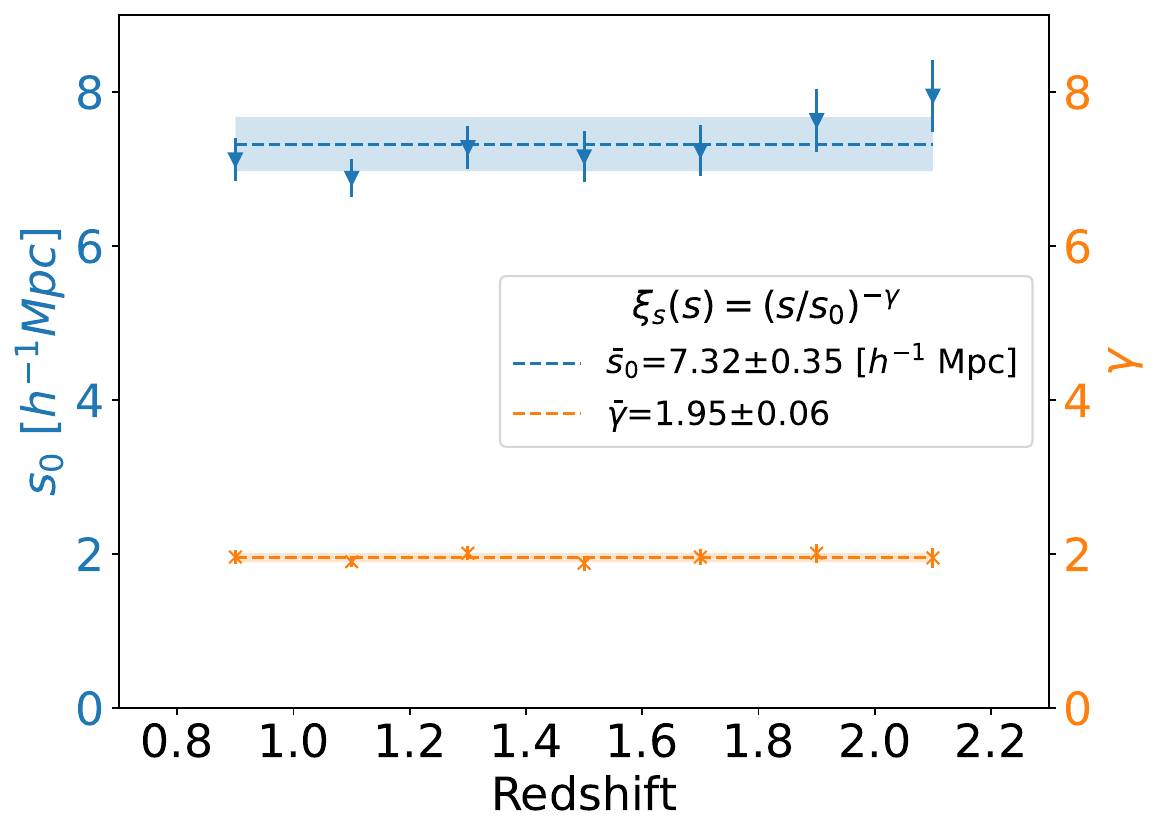}
    \caption{The best-fit parameters of the auto-correlation TPCF within each redshift bin, for the entire eBOSS quasar sample in our target field. The blue triangles at the top show the best-fit values for the correlation length $s_0/h^{-1}\ \mathrm{Mpc}$, while the orange crosses at the bottom show the best-fit values for the slope $\gamma$, assuming no priors in the fit. The dashed lines trace the average values for each parameter, with the uncertainties marked by the shaded areas.}
    \label{fig:clustering_amp_full}
\end{figure}

In this section, we start by testing our derived redshift space correlation length ($s_0$) for the entire sample and compare our result with previous work on quasar clustering \citep{ross_clustering_2009,white_clustering_2012,eftekharzadeh_clustering_2015,chaussidon_angular_2022}, while in Section~\ref{sec:bias} we discuss the quasar bias ($b$) calculated from our correlation length measurements. As shown in Figure~\ref{fig:opticalproperty}, we divided our sample into seven redshift bins from $0.8<z<2.2$, each covering a range of $\Delta z=0.2$. In Figure~\ref{fig:clustering_amp_full}, we present the real space correlation lengths of the full quasar sample in each of the redshift bins, following the method described in Section~\ref{sec:tpcf}. Note that, contrary to the rest of the fits conducted in this work, no priors have been used in deriving this fit. The best-fit values of $\gamma$ remain largely unchanged throughout the redshift range with a redshift-averaged value (denoted $\bar\gamma$) of $\bar \gamma=1.95\pm0.06$, while the best-fit values of $s_0$ show indications of a weak redshift evolutionary trend (see Table~\ref{tab:result_1}), where a larger redshift sees a higher correlation length up to $s_0=7.95\pm0.47\ h^{-1}\ \textrm{Mpc}$ at $2.0<z<2.2$, but the differences are within the uncertainty range. The averaged correlation length across the redshift bins is $\bar s_0=7.32\pm0.35\ h^{-1}\ \textrm{Mpc}$, which is in broad agreement with the literature results in the works listed above, where different values range from $6-10\ h^{-1}\ \textrm{Mpc}$ for quasars in a similar redshift space \citep[see Figure~12 in][]{retana-montenegro_probing_2017}.

\subsection{Priors and additional constraints}
\label{sec:prior}

Due to redshift space distortions and statistical fluctuation caused by the sparsity of quasar samples, the TPCF will not be a perfect power law across the entire distance scale, and the best-fit parameters will depend on the assumed priors and ranges of distance scale within which we fit the TPCF. \par 

In our analysis, we apply a fixed prior to the power-law slope ($\gamma=1.95$) based on the best-fit result shown in Figure~\ref{fig:clustering_amp_full}, since the data quality of the high-power jet-dominated quasar sub-sample does not allow a flexible fit of the power-law slope, and we need to maintain consistency in the comparison between the different sub-populations of quasars. This is broadly in agreement with the values used in the previous literature \citep[e.g. $\gamma=2.0$;][and references therein]{shen_quasar_2009}. \par

For non-jet-dominated (SF and RI) quasars, we use a fitting range of $[s_1,s_2]=[5,100]h^{-1}\ \textrm{Mpc}$ for the correlation length, which is in agreement with the fitting range used in \citet{shen_quasar_2009} and \citet{retana-montenegro_probing_2017}. We adopted a fitting range of $[s_1,s_2]=[10,200]h^{-1}\ \textrm{Mpc}$ for jet-dominated quasars (lo-jet and hi-jet) since for these populations, significant correlation signals still remain at larger distance scales. The lower distance scale limit of the fitting range is motivated by three factors that affect the TPCF at small distance scales: (a) the redshift space distortion effect, which flattens the TPCF on small scales; (b) the one-halo term (correlation between quasars in the same halo) which is expected to be small but can boost the TPCF above the single-power law fit; (c) the SDSS fibre collision constraints, which can decrease the number of close pairs in the data, although this should be largely corrected by weights of the data and random samples. Meanwhile, the upper limit of the fitting range is driven by both the signal-to-noise limits of the data and the steepening of TPCF at large distance scales, also as a result of the redshift space distortion effect. \par

We examine the choice of our fitting ranges by upscaling the minimum distance ($5\ h^{-1}\ \textrm{Mpc}$) to $10$ and $15\ h^{-1}\ \textrm{Mpc}$, and downscaling the maximum distance ($200\ h^{-1}\ \textrm{Mpc}$) to $150$ and $100\ h^{-1}\ \textrm{Mpc}$. The differences from the best-fit correlation lengths within different fitting ranges are within the uncertainties of the original fit, indicating that our results are not impacted by the choice of fitting ranges. We therefore picked the current fitting ranges since they give the highest signal-to-noise ratios. \par

Occasionally, the values of TPCF in some distance scales fall below zero due to statistical fluctuations or systematics, which compromises the quality of chi-square fit since the fit is performed in log-log space. To mitigate these negative values in the actual fit, we assign a sufficiently small value ($10^{-10}$) to these data points, with its uncertainty covering the upper and lower limits of the original data points. However, having too many negative $\xi_s$ measurements in the fitting range suggests that the assumption of a power-law TPCF may have broken down, possibly due to low number statistics. To keep our measurements robust, if more than 1/3 of the $\xi_s$ measurements are negative when fitting the TPCF for a certain quasar population at any given redshift bin, we switch to a set of wider redshift bins (see descriptions in specific sections), so that the fitting of TPCF is always performed with predominantly positive data points. \par

\subsection{Evolution with relative jet power}

In this section, we present the evolution of quasar clustering with their relative jet strengths (fraction of jet in their radio emission, characterised by the four populations defined in Section~\ref{sec:model}) and the redshift. \par

\begin{figure}
    \centering
    \includegraphics[width=1\linewidth]{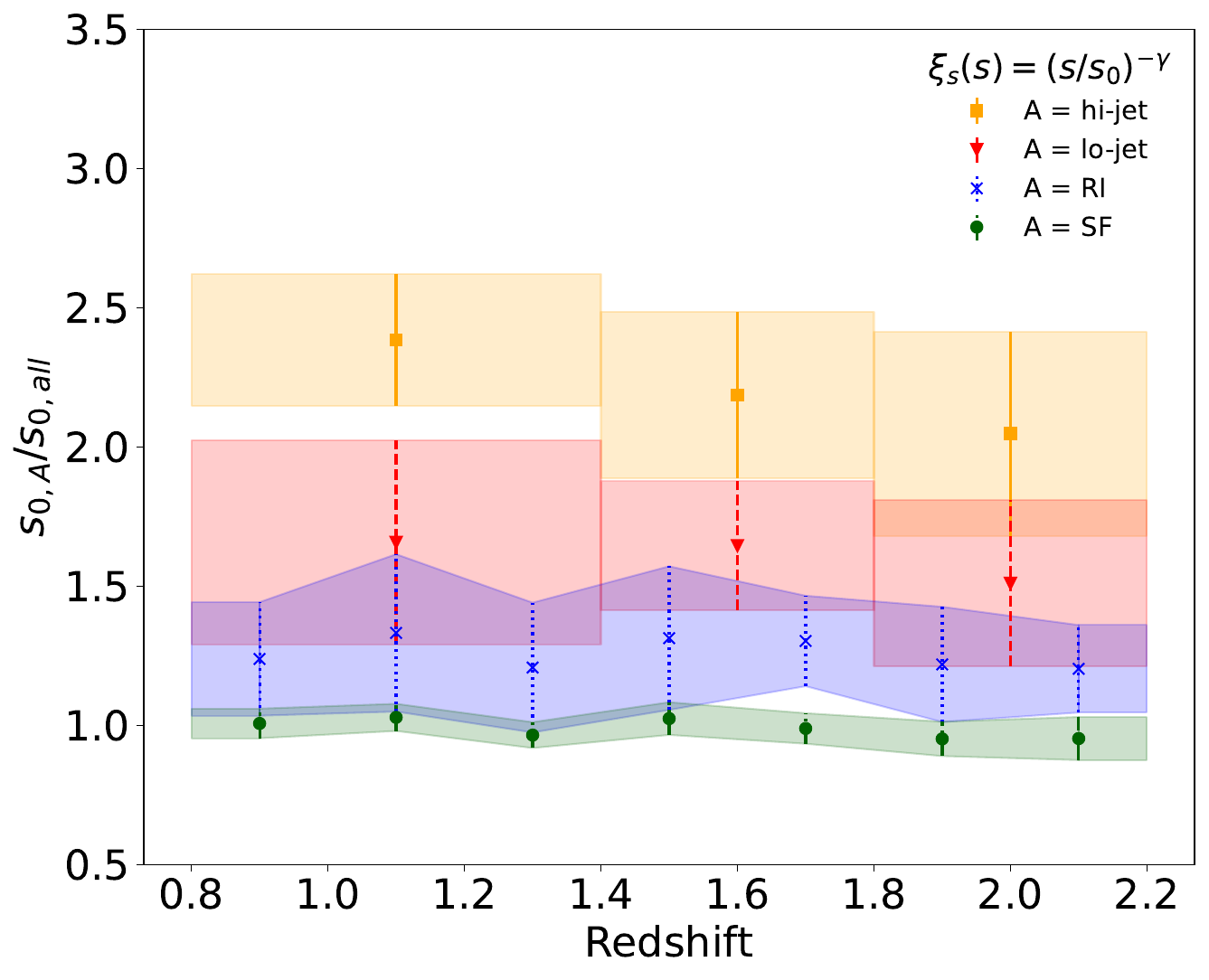}
    \caption{The normalised correlation lengths ($s_{0,A}/s_\mathrm{0,all}$) for each quasar population defined in this work, within each redshift bin. $s_\mathrm{0,all}$ are the correlation lengths of the entire quasar sample in the corresponding redshift bin, and $s_{0,A}$ are the correlation lengths of population $A$ (SFQ for the green solid dots, RIQ for the blue crosses, lo-jet quasars for the red inverted triangles, hi-jet quasars for the orange squares) within the same redshift bin. The shaded areas highlight the uncertainty levels of each measurement.}
    \label{fig:clustering_amp_ratio_sf_agn}
\end{figure}

Figure~\ref{fig:clustering_amp_ratio_sf_agn} compares the trends of normalised correlation lengths ($s_{0,A}/s_\mathrm{0,all}$; A stands for different quasar sub-populations) of the SFQ, RIQ, lo-jet, and hi-jet quasars, using the correlation length of the entire sample ($s_\mathrm{0,all}$) as a normalisation factor. Note that due to constraints in number statistics, for lo-jet and hi-jet quasars we fit the corresponding TPCFs within redshift bins of $0.8<z<1.4$, $1.4<z<1.8$, and $1.8<z<2.2$, and $s_\mathrm{0,all}$ is determined by the best-fit correlation lengths at bin-centre redshifts. We observe a clear trend in which quasars with a higher fraction of jet contribution are more clustered in all of the redshift bins, with no significant redshift dependence. The hi-jet quasar population is always more clustered than other populations in all redshift slices, with correlation lengths $\sim2.5$ times higher than those of the SFQ population. \par

Table~\ref{tab:result_1} lists the redshift space correlation lengths ($s_0$) of the four radio quasar populations across redshift coverage ($0.8<z<2.2$). In all redshift bins, $s_{0,\textrm{hi-jet}}$ is considerably larger than $s_{0,\textrm{SFQ}}$, with an average value of $s_{0,\textrm{hi-jet}}=16.0\pm 1.2\ h^{-1}\ \textrm{Mpc}$. SFQs have an average correlation length of $s_{0,\textrm{SFQ}}=7.23\pm 0.19\ h^{-1}\ \textrm{Mpc}$, which is in good agreement with the results on RQ quasars in \citet{shen_quasar_2009} and \citet{retana-montenegro_probing_2017}. Our hi-jet samples show a higher correlation length than the RL quasar samples in the previous literature: $r_0=12.97\pm 2.47\ h^{-1}\ \textrm{Mpc}$ at $0.4<z<2.5$ for FIRST-autocorrelated quasars in \citet{shen_quasar_2009}. This is possibly due to the model-motivated selection being less contaminated by sources with weaker jets compared to their total radio emission, and the selection criterion of this population is more strict than the traditional RL quasars; in other words, they have a higher relative jet power. The average correlation lengths for the RIQs and lo-jet quasars are $s_{0,\textrm{RIQ}}=9.4\pm0.6\ h^{-1}\ \textrm{Mpc}$ and $s_{0,\textrm{lo-jet}}=11.7\pm1.2\ h^{-1}\ \textrm{Mpc}$, respectively. The measured correlation lengths of the lo-jet quasar population are consistent with those of the FIRST-detected quasars \citep{shen_quasar_2009,retana-montenegro_probing_2017}, which is expected since the lo-jet population makes up the majority of the traditional RL quasar samples. \par

\subsection{Black hole mass dependence}
\label{sec:result_mass}

Evidence from observations and simulations suggests that the BH mass is correlated with both the halo mass \citep[e.g.][]{ferrarese_beyond_2002,bandara_relationship_2009,booth_dark_2010} and the quasar radio emission \citep[e.g.][]{mclure_relationship_2004,yue_novel_2025}. It is therefore important to break the degeneracy between halo mass and BH mass: are quasars with higher jet fraction more clustered because they indeed reside in more massive haloes, or is it because they host more massive BHs that tend to be found in more massive haloes? While the previous sections investigated the correlation between halo mass and quasar radio emission, here we will also investigate the influence of the BH mass.  \par

\begin{figure}
    \centering
    \includegraphics[width=1\linewidth]{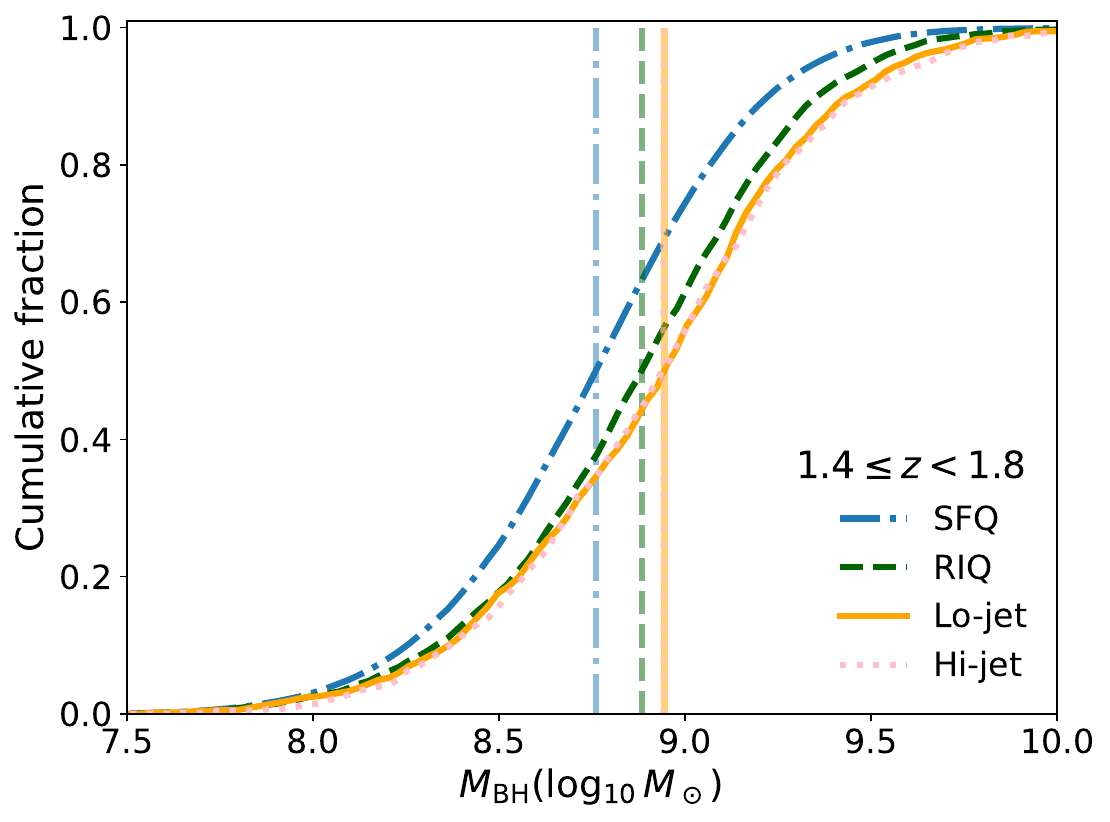}
    \caption{Cumulative distributions of black hole masses in SFQ (dash-dotted), RIQ (dashed), lo-jet (solid) and hi-jet (dotted) quasar populations in a representative redshift slice of $1.4<z<1.8$. The vertical lines mark the median BH masses of each populations.}
    \label{fig:bhmass_jetfrac}
\end{figure}

Firstly, we do not find a characteristic BH mass range for any of the quasar populations defined in this work, from SFQs to hi-jet quasars. Figure~\ref{fig:bhmass_jetfrac} shows the BH mass distribution in one representative redshift slice ($1.4<z<1.8$) used in our clustering analysis. Although the average BH masses show a mild evolutionary trend with jet fraction, the overall distributions of BH masses span the same range.\par

\citet{yue_novel_2025} found that when examining the effect of BH mass on quasar radio emission, a difference in the typical radio jet powers exists only in the top 20\% of the BH mass distribution within a $\mathcal{M}_i-z$ grid cell, which explains the differences in average BH masses across different populations. To investigate the influence of this effect on our clustering measurements, we divide our quasar sample into four different sub-populations based on their jet fraction and BH masses: \par

\begin{enumerate}
    \item Quasars hosting moderate mass BHs (the lower 80\% in the BH mass distribution of quasars within the corresponding $\mathcal{M}_i-z$ grid cell) with radio emission dominated by SF activity;
    \item Quasars hosting massive BHs (top 20\% in the BH mass distribution of quasars within the corresponding $\mathcal{M}_i-z$ grid cell) with radio emission dominated by SF activity;
    \item Quasars hosting moderate mass BHs with radio emission dominated by AGN jet activity;
    \item Quasars hosting massive BHs with radio emission dominated by AGN jet activity.
\end{enumerate}

Note that by classifying within the same $\mathcal{M}_i-z$ grid cells, quasars across the four sub-populations share similar optical luminosities and redshifts, i.e., have similar distributions in $\mathcal{M}_i-z$ space. To obtain the statistics required for clustering analysis, we combined the lo-jet and hi-jet quasar populations into one jet-dominated quasar population, where the AGN jet still dominates the radio emission in this quasar population. We compare these quasars against the SFQ population. Due to constraints in number statistics, for jet-dominated quasars we will only calculate the TPCFs within two redshift bins: $0.8<z<1.6$ and $1.6<z<2.2$. In both jet-dominated and SFQ samples we exclude sources that have spurious BH mass measurements ($M_\mathrm{BH}\leq0$) in the SDSS DR16Q catalogue and are therefore not included in the classification. By quoting the lower limit (5-th percentile), upper limit (95-th percentile), and median values, Table~\ref{tab:bhmass} describes the BH mass distributions of quasar populations split into four sub-populations, within the two redshift bins explored in this section. Quasars with moderate BH masses have similar BH mass distributions spanning a wide range up to $M_\mathrm{BH}\sim10^9M_\odot$. Quasars with top 20\% BH masses in each grid cell host BHs with $M_\textrm{BH}>10^{8.5}M_\odot$, also covering the same range, although jet-dominated quasars host more massive BHs than SF-dominated quasars on average. \par

\begin{table}
    \centering
    \begin{tabular}{lcc}\toprule
           $\log (M_\mathrm{BH}/M_\odot)$ &Moderate $M_\textrm{BH}$ &High $M_\textrm{BH}$\\\midrule
           &\multicolumn{2}{c}{$0.8<z<1.6$}\\
           Least 5\% &$7.82\ (7.92)$ &$8.65\ (8.76)$ \\
           Median &$8.48^{+0.30}_{-0.38}\ (8.58^{+0.31}_{-0.38})$ & $9.03^{+0.25}_{-0.21}\ (9.17^{+0.27}_{-0.25})$ \\ 
           Maximum 95\% &$8.96\ (9.05)$ &$9.45\ (9.69)$ \\ \midrule
           & \multicolumn{2}{c}{$1.6<z<2.2$}\\
           Least 5\% &$8.02\ (8.00)$ &$8.97\ (9.00)$ \\
           Median &$8.72^{+0.30}_{-0.39}\ (8.78^{+0.31}_{-0.41})$ & $9.25^{+0.24}_{-0.19}\ (9.36^{+0.26}_{-0.21})$ \\ 
           Maximum 95\% &$9.18\ (9.26)$ &$9.66\ (9.84)$ \\ \bottomrule
    \end{tabular}
    \caption{Characteristic points in the BH mass distributions of quasars hosting moderate-mass BHs and high-mass BHs and with different dominant sources of radio emission, across the two redshift bins investigated in this work ($0.8<z<1.6$ and $1.6<z<2.2$). The first value in each grid cell represents the SF-dominated quasars, while the second (bracketed) value represents the jet-dominated quasars. The uncertainties in the median BH masses are given by the 16-th and 84-th percentiles of the mass distributions. All values are in log units ($\log (M_\mathrm{BH}/M_\odot)$).}
    \label{tab:bhmass}
\end{table}

\begin{figure}
    \centering
    \includegraphics[width=1\linewidth]{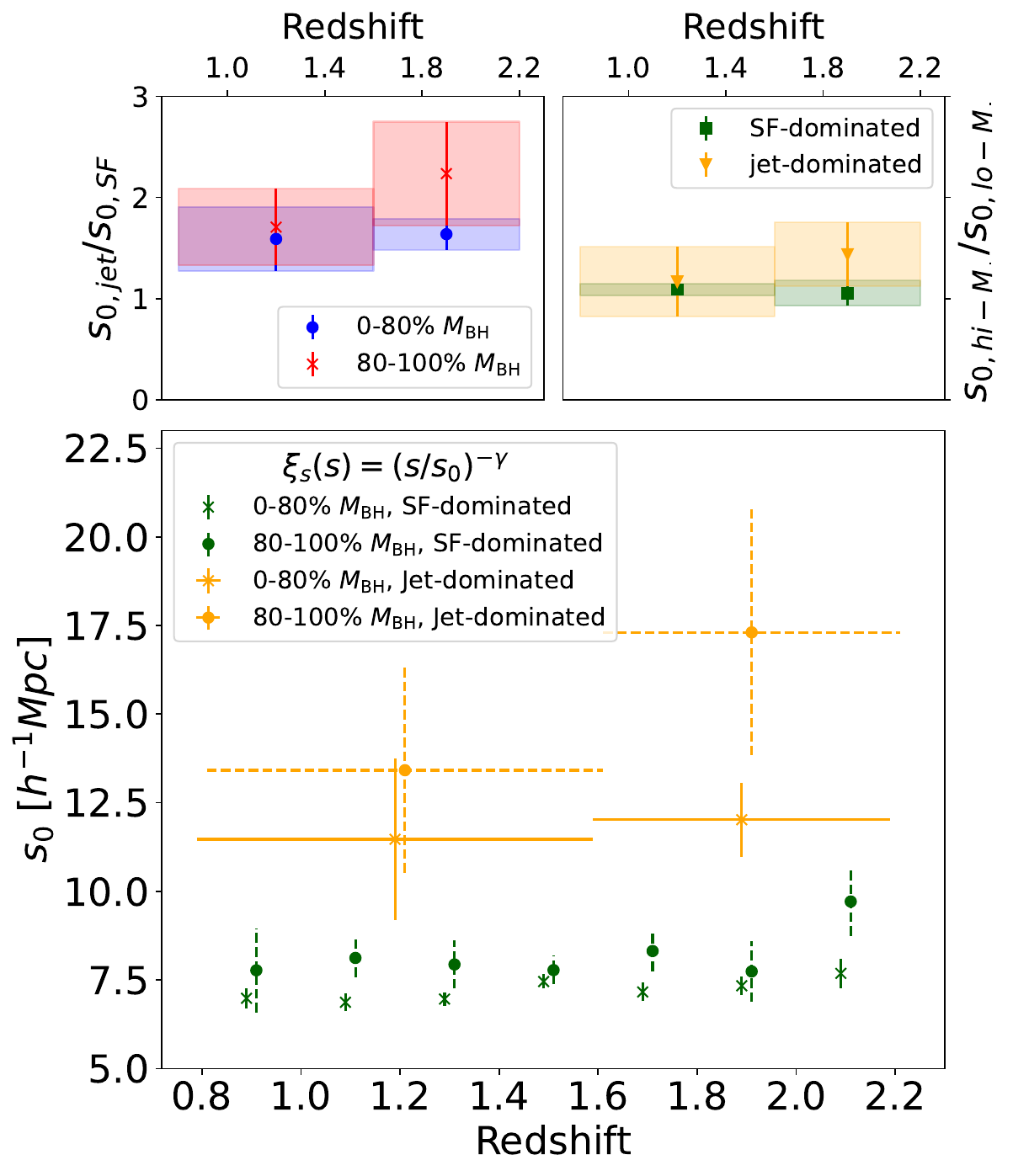}
    \caption{Lower: the correlation lengths ($s_0$) measured from auto-correlation functions of four quasar sub-populations split by BH mass percentiles and dominating radio sources: crosses stand for moderate BH mass ($0-80$ percentile within each $\mathcal{M}_i-z$ bin); solid dots  stand for high BH mass ($80-100$ percentile); dark green colour marks the SF-dominated population; and bright orange colour marks jet-dominated quasars (combining lo-jet and hi-jet quasar populations). The uncertainties in correlation lengths measurements are shown in solid or dashed error bars along the y-axis. For jet-dominated quasars, the redshift bin sizes are shown with the error bars along the x-axis. Data points with the same redshifts are shifted by a small amount along the x-axis to improve clarity. Upper: The ratio between correlation lengths measurements from quasars with different dominant radio source (left) and different BH mass (right), while controlling the other parameter. The uncertainty ranges are shown by the shaded areas. The differences in the correlation lengths between jet-dominated and SF-dominated quasars are larger than those between quasars with high BH masses and intermediate BH masses, indicating that BH mass is not the primary force driving the evolutionary pattern in the lower panel.}
    \label{fig:clustering_amp_quads}
\end{figure}

The main (lower) panel of Figure~\ref{fig:clustering_amp_quads} compares the correlation lengths of the four quasar sub-populations, where round data points show populations with moderate BH mass, and crossed data points track populations with high BH mass. The dark green points trace the SF-dominated quasars, while bright orange points trace the jet-dominated quasars. Note that for a clearer presentation, the data points have been shifted by a small amount in redshift. Quasar populations with the top 20\% most massive BHs are indeed more clustered than quasars with moderate mass BHs (within the same $\mathcal{M}_i-z$ grid), both in SF-dominated quasars and jet-dominated quasars, but the difference introduced by different BH masses is far smaller than the difference between SFQs and jet-dominated quasars. In particular, SF-dominated quasars with high BH mass are still less clustered than jet-dominated quasars with moderate BH mass at similar redshifts and optical luminosities. \par

In the two upper sub-panels of Figure~\ref{fig:clustering_amp_quads}, we further compare the relative difference in correlation lengths assuming the same BH mass percentiles (upper left) or the same dominant radio source (jet or SF; upper right), respectively. When controlling for the BH mass percentile, the correlation lengths of jet-dominated quasars are 1.5-2.5 times larger than the correlation lengths of SFQs. In contrast, the differences between the correlation lengths of high-BH-mass quasars and moderate-BH-mass quasars when controlling for jet fraction are between a factor of 1 and 1.5. There are signs of a weak redshift evolution in the jet-dominated quasar population, where the BH mass impact on the correlation length becomes more profound in higher redshifts. If real, it could be due to the higher merger fraction at cosmic noon, which leads to more jet activity being triggered in denser environments created by merger processes that link to larger BH mass. However, constraining such a trend with current data is hard, and this may also be driven by a selection effect, where less massive jet-dominated quasars are harder to detect at larger redshifts. This bias would lead to an underestimation of the clustering signal; therefore more extensive data are required before drawing a conclusion.

\subsection{Quasar luminosity dependence}

Quasar bolometric luminosity scales with BH mass and the accretion rate, and is reported to have a weak correlation with halo mass through clustering measurements \citep[e.g.][]{shen_quasar_2009}; however, such a difference only exists in the most luminous objects while no general trend between halo mass and quasar luminosity has been found in recent studies \citep{krolewski_measuring_2015,he_clustering_2018}. Since our quasar populations are stacked across different luminosity-redshift bins, any luminosity dependence would be expected to have a minor effect on the clustering properties. \par

To test for this luminosity dependence, within each redshift slice, we split our quasars into the brightest 50\% and the faintest 50\% populations based on their $\mathcal{M}_i$ distribution. This is to maximise the statistics for both the bright and faint quasars within each bin. We also compare the top 10\% brightest quasars with the rest of the population to study the extreme population in our sample. We then measured the TPCFs for these quasar sub-populations in each redshift slice. \par

\begin{figure}
    \centering
    \includegraphics[width=1\linewidth]{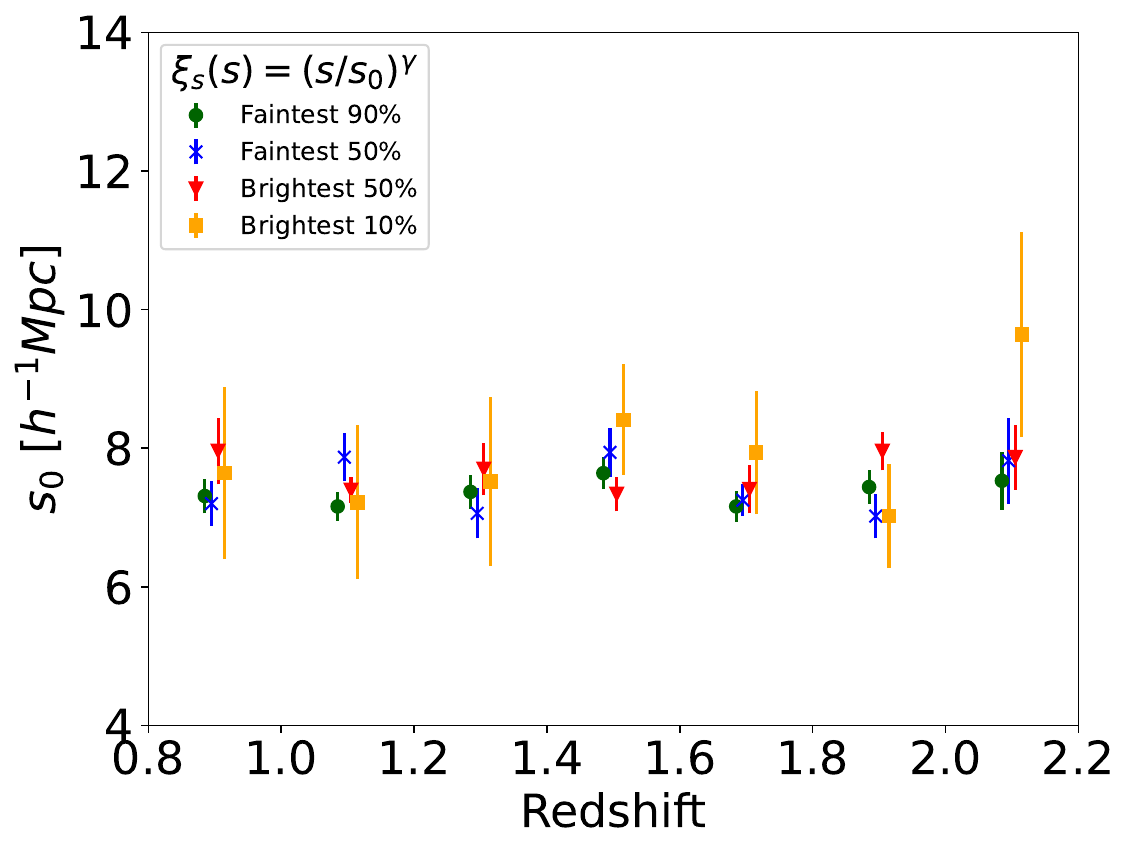}
    \caption{The correlation lengths of quasar sub-populations stacked across different redshift slices with widths $\Delta z=0.2$, defined with different \emph{i}-band luminosity percentiles in each slice. Quasars are split with two divides in the \emph{i}-band luminosity: between the brighter 50\% (red inverted triangles) and the fainter 50\% (blue crosses) divided at the median of the distribution, and between the top 10\% most luminous quasars (orange squares) and the rest of the population (dark green dots). The uncertainty ranges of correlation length measurements are marked with the error bars along the y-axis. Measurements at the same redshifts have been shifted by a small amount along the x-axis to improve clarity.}
    \label{fig:clustering_amp_lum}
\end{figure}

Figure~\ref{fig:clustering_amp_lum} shows the correlation lengths of quasar populations with different luminosities defined above. In all redshift slices, we see no significant differences between the correlation lengths of the brighter and fainter halves of the luminosity distribution. At some redshifts, there is a weak suggestion that the most luminous 10\% of quasars are marginally more clustered, but differences are small and of low significance, compared to the scale of differences between the jet- and SF-dominated quasar populations in Figure~\ref{fig:clustering_amp_quads}. We conclude that there is no luminosity selection effect or luminosity dependence driving the differences in clustering measurements that we see in our quasar sample.

\section{Quasar bias and host halo mass}
\label{sec:bias}

Based on predictions of the linear perturbation theory of structure formation, quasars are biased tracers of the dark-matter haloes that they reside in \citep[e.g.][]{springel_simulations_2005}. The linear bias factor, which describes how quasars spatially cluster according to the underlying dark matter, is typically defined as \citep[e.g.][]{peacock_halo_2000}: 

\begin{equation}
    b^2=\frac{\xi_\textrm{QSO}(r,z)}{\xi_\textrm{DM}(r,z)},
\end{equation}

\noindent where $\xi_\textrm{QSO}$ and $\xi_\textrm{DM}$ are the quasar and dark matter correlation function \citep{peebles_large-scale_1980}. Since the evolution of the bias factor with redshift is scale independent \citep{peebles_large-scale_1980}, we can therefore calculate the bias factor at $s=8\ h^{-1}\textrm{Mpc}$, where $\xi_\textrm{QSO}(8,z)=\left[s_0(z)/8\right]^{-\gamma}$, and the dark matter correlation function given by \citet{peebles_large-scale_1980} is $\xi_\textrm{DM}=\sigma_8^2(z)/J_2$ \citep[see e.g.][]{lindsay_galaxy_2014,hale_clustering_2018}. Here, $J_2=72/[(3-\gamma)(4-\gamma)(6-\gamma)2^\gamma]$, and the parameter $\sigma_8^2$ is the dark matter density variance in a comoving sphere with a radius of $r=8\ h^{-1}\textrm{Mpc}$. The scale-independent evolution of the bias factor with redshift can thus be calculated given the correlation length $s_0(z)$ using:

\begin{equation}
    b(z)=\left(\frac{s_0(z)}{8}\right)^{-\gamma/2}\frac{J_2^{1/2}}{\sigma_8D(z)/D(0)}.
\end{equation}

\noindent Here $D(z)$ is the growth factor at redshift $z$ \citep{carroll_cosmological_1992} and is calculated with the equation in \citet{hamilton_formulae_2001}. With the bias factor, one can derive the typical halo mass of the given quasar population from different numerical halo collapse models of choice, including \citet{jing_accurate_1998}, \citet{sheth_ellipsoidal_2001}, and \citet{tinker_large-scale_2010}. \par

Deriving the quasar bias and host halo mass from the correlation length measured from the TPCF relies on many assumptions in the halo assembly history and the halo occupation model, which are not always correct. Due to the limited statistics in the jet-dominated quasar population, in this section we will only make \emph{indicative} (order-of-magnitude) estimations on the quasar bias and halo mass, based on the assumptions that quasars are accurate tracers of the baryon distribution within a halo and that dark matter haloes above a certain mass are capable of hosting only one quasar, which resides at the centre of the halo. In other words, we only consider the two-halo term of the TPCF and ignore the one-halo term which traces the pairs of galaxies residing in the same halo, which is a sensible assumption to make in the clustering analysis of quasars or radio galaxies \citep[e.g.][]{magliocchetti_2df_2004,shen_quasar_2009}. We also assume a numerical halo collapse model proposed by \citet{tinker_large-scale_2010}. \par

\begin{figure}
    \centering
    \includegraphics[width=1\linewidth]{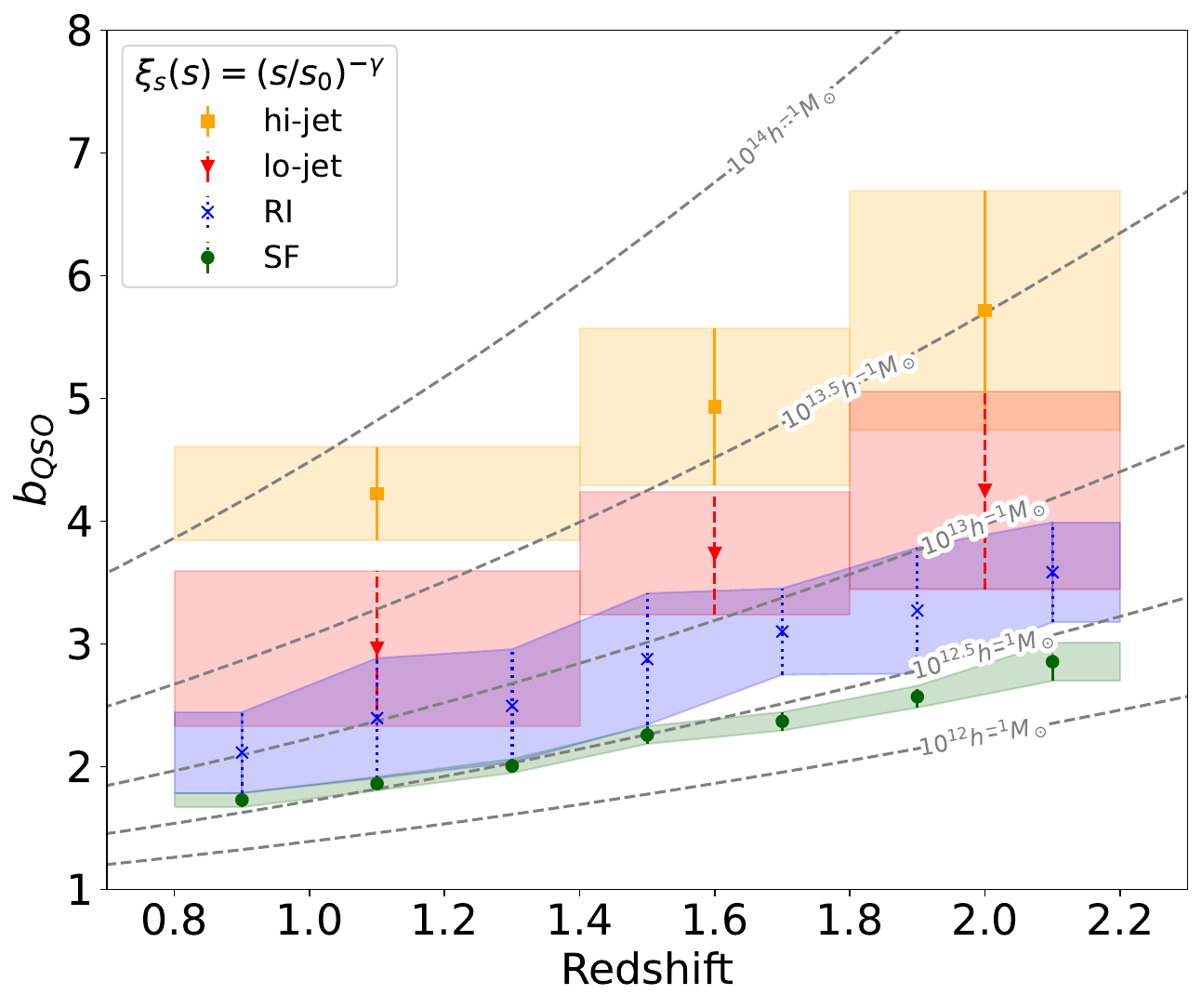}
    \caption{Quasar bias of SFQ (dark green dots), RIQ (blue crosses), lo-jet (red inverted triangles), and hi-jet (orange squares) quasar populations and their evolution with redshift, measured from the two-halo term in auto-correlation TPCFs. The shaded areas mark the uncertainties in bias measurements, where the lo wer and upper limit are values computed with the 16-th and 84-th percentile in correlation length measurements, respectively. The grey dashed lines mark the theoretical bias evolution of haloes with masses of $10^{12}h^{-1}\ M_\odot$, $10^{12.5}h^{-1}\ M_\odot$, $10^{13}h^{-1}\ M_\odot$, $10^{13.5}h^{-1}\ M_\odot$, $10^{14}h^{-1}\ M_\odot$, respectively (from bottom to top), assuming the relationship between bias and halo mass in the numerical model from \citet{tinker_large-scale_2010}.}
    \label{fig:clustering_bias_sf_agn}
\end{figure}

Figure~\ref{fig:clustering_bias_sf_agn} shows the quasar bias factors calculated for the SFQ, RIQ, lo-jet, and hi-jet quasar populations, while the dashed lines show the redshift evolution of the bias factor $b(\nu)$, where $\nu$ is the peak height of the power spectrum and evolves with redshift through $\nu(z)\propto1/D(z)$; $b(\nu)$ is computed with different underlying halo masses (from bottom to top: $10^{12}$, $10^{12.5}$, $10^{13}$, $10^{13.5}$, $10^{14}h^{-1}\ M_\odot$), assuming the \citet{tinker_large-scale_2010} numerical model and the numerical conversion relationship between the halo mass and the peak height $\nu$ in \citet{seppi_mass_2021}. The bias factors for the SFQs show an evolutionary trend with redshift that is consistent with a typical halo mass of $10^{12.5}h^{-1}\ M_\odot$. This is also consistent with the typical halo mass of RQQs both in observations \citep{retana-montenegro_probing_2017} and simulations \citep{wilman_semi-empirical_2008}. In line with the increasing values of $s_0$, the bias factors increase as each quasar population becomes increasingly dominated by jet activities, again indicating that the jet fraction is \emph{positively correlated} with the halo mass. At $z<1.4$, the quasars that host the most powerful radio jets (hi-jet quasars) tend to reside in rich cluster environments with typical halo masses of $10^{13.5-14}h^{-1}\ M_\odot$, and for lo-jet quasars a typical halo mass of $10^{13-13.5}h^{-1}\ M_\odot$. This is in agreement with the typical halo mass calculated in \citet{petter_environments_2024} where they investigated a subsample of radio galaxies detected in the LoTSS wide- and deep-tier (with B\"ootes field only) survey with radio luminosities $L_\mathrm{144MHz}>10^{25.5}\mathrm{W}\cdot\mathrm{Hz}^{-1}$. Their radio luminosity range largely overlaps with the traditional radio-loud quasar population and the jet-dominated quasar populations defined in this work; however, their sample also includes a large population of LERGs. We do not see a significant evolutionary trend between the characteristic halo masses and redshift in SFQ and RIQ populations, while for lo-jet and hi-jet quasars there is a mild trend of halo mass decreasing towards higher redshifts at $z<2$; this is consistent with the predictions of the \citet{fakhouri_merger_2010} halo growth model based on a progenitor halo mass of $\sim10^{13}h^{-1}\ M_\odot$ at $z=2.5$ \citep[see also the discussion in][]{petter_environments_2024}. However, uncertainties in these measurements are large and detailed halo occupation distribution (HOD) modelling is required to either reliably demonstrate or rule out the redshift evolution.

\section{Linking halo environment to jet powering mechanisms}
\label{sec:discussion}

When we investigate the difference in the large-scale clustering between different radio quasar populations, we find evidence to the possible mechanisms that create such a difference in the radio jet power between these populations. In other words, our aim is to answer the question: what role does the large-scale environment play in determining the power of radio jets hosted by quasars? \par
 
Before we start, we summarise some of the key findings from the clustering measurements in the previous sections:

\begin{enumerate}
    \item Correlation lengths measured from TPCF increase monotonically with jet fraction, from the SF to the hi-jet quasar population, without any bimodality (i.e., jet-dominated quasars still cluster differently according to their jet fraction, and RIQs cluster differently from both SFQs and jet-dominated quasars);
    \item Quasars with a higher jet fraction and moderate BH mass are more clustered than quasars with a lower jet fraction but higher BH mass; this indicates that the relationship between halo mass and jet power is fundamental, and not solely a secondary effect due to both properties correlating with black hole mass or bolometric luminosity;
    \item Jet-dominated quasar populations have characteristic halo masses of rich cluster environments ($\sim10^{13-14}h^{-1}\ M_\odot$), which is significantly more than an order of magnitude higher than the halo masses of typical SDSS quasars \citep[$\sim10^{12}h^{-1}\ M_\odot$; e.g.][]{ross_clustering_2009};
    \item  There is no significant redshift evolutionary trend in the characteristic halo masses of the SF-dominated quasar population, while there are signs of a mild redshift evolution in the halo masses of jet-dominated quasar populations, consistent with the halo growth model.
\end{enumerate}

We first discuss the possibility where the increase in the jet fraction in more massive haloes may be the result of a higher \emph{probability} of radio jets being triggered in denser environments on Mpc scales \citep[e.g.][]{hatch_why_2014}. While this may be true, we found that the correlation length continues to increase from the lo-jet to the hi-jet regime, where both populations are dominated by powerful jets, suggesting that halo mass does not \emph{only} affect the triggering of jets, and therefore the halo mass needs to be responsible for the \emph{powering} of radio jets. \par

Based on these results, one can easily conclude that jet power is positively correlated with large-scale clustering, hence the halo mass. However, multiple mechanisms may still be at play to establish what seems like a simple correlation; as we try to discuss them below. \par

\subsection{Radio luminosity boost by lobe confinement}

The first thing to consider is whether there are any observational systematics of overestimating jet power by measuring its radio luminosity. Simulation results \citep[e.g.][]{hardcastle_radio_2020} show that the observed radio luminosity in jet-dominated AGN correlates well with the jet mechanical power; however, the interaction between the jet and the surrounding medium may cause a significant ($>1\ \mathrm{dex}$) scatter to this correlation, as revealed in simulations of the radio jet dynamics \citep[][]{hardcastle_numerical_2013,hardcastle_simulation-based_2018}. When the radio jet propagates into a dense environment, it will suffer from decollimation and depolarisation, resulting in slower jet speeds. The radio lobe will therefore be confined within a smaller spatial scale, and the energy loss from adiabatic expansion will be reduced. As a result, the observed radio source can be brighter in denser environments (i.e. massive clusters/haloes) despite little difference in the actual jet power. Observationally, this has been confirmed in resolved studies of several nearby objects \citep[e.g.][]{barthel_anomalous_1996}; simulations in \citet{hardcastle_simulation-based_2018} also show that within a narrow redshift range ($z<0.5$), the scatter in the observed relationship between the 150 MHz radio luminosity ($L_{150}$) and the jet mechanical power ($Q$) could be mostly driven by environmental factors characterised by the cluster mass, where sources residing in more massive clusters typically have higher $L_{150}$ given fixed $Q$. \par

However, these differences are relatively small compared to the range of potential jet powers of radio sources, which span several orders of magnitudes. In the simulated samples of \citet{hardcastle_simulation-based_2018}, the observed radio luminosity itself has little correlation with the cluster mass; for a fixed cluster mass, the variation in radio luminosity $L_{150}$ is mainly driven by the variation in $Q$ instead of the scatter within the $L_{150}-Q$ correlation. In contrast, we observe a significant difference between jet luminosity and quasar clustering in our jet-dominated quasar populations, which is in line with the local correlations between radio luminosity and cluster richness measured from X-ray luminosity \citep{ineson_radio-loud_2013,ineson_link_2015}. The contrast between the results from the simulated sample and the observed sample originates from the assumption in the simulation that a jet of any power can be switched on in any environment. This suggests that the environmental boost to the observed radio luminosity at fixed jet power plays a minor role in the observed difference and that some \emph{intrinsic correlation} between jet power (or jet lifetime) and environment richness must be assumed to explain the differences in observation. \par

\subsection{Variations in accretion mode}

Popular theories of radio jet production suggest that the jet power is determined by the magnetic field extracting energy either directly from the accretion flow \citep[B-P model;][]{blandford_hydromagnetic_1982} or from the black hole spin \citep[B-Z model;][]{blandford_electromagnetic_1977}; we will discuss these two factors in this and the following sections. 

In the B-P mechanism, centrifugal forces associated with the magnetic field transfer the angular momentum of the accretion flow out to collimated jets along the magnetic field lines. The jet power is therefore determined by the structures of the magnetic field and the accretion disc, with a higher jet production efficiency typically found in hot accretion modes where the disc has a thicker geometry \citep[see e.g. the review by][]{yuan_hot_2014}. However, since 95\% of the SDSS DR16 quasars have Eddington ratios between $0.01\leq\lambda_\mathrm{Edd}\equiv (L_\mathrm{bol}/L_\mathrm{Edd})\leq1$ \citep[see Figure 2 in][]{wu_catalog_2022}, the jet power in our quasar sample is unlikely to be directly correlated with the accretion efficiency. Moreover, there is no major difference between the typical halo masses of jet-dominated \emph{quasars} derived in this study, which are mostly made up of HERGs, and the halo masses of radio \emph{AGN} \citep[e.g.][]{petter_environments_2024}, in which LERGs \citep[with typical Eddington ratios $\lambda_\textrm{Edd}<0.01$; see][]{heckman_coevolution_2014} are the dominant population. Both populations tend to reside in cluster environments with halo masses $\sim10^{13-14}\ h^{-1}M_\odot$. We note that \citet{hale_clustering_2018} reports a difference in the typical halo masses of high-to-moderate-luminosity AGN (HLAGN, equivalent to HERGs) and moderate-to-low-luminosity AGN (MLAGN, equivalent to LERGs), where HLAGN typically occupy haloes with masses of $M\sim10^{13}\ h^{-1}M_\odot$, while for MLAGN the typical halo masses are $M\sim10^{13.5}\ h^{-1}M_\odot$. However, these values are within the wide ranges of halo masses determined in this work and in \citet{petter_environments_2024}. These values are also consistently higher than the typical halo masses of SFQs and RIQs derived in this work. To summarise the discussion above: we found rich halo environments are efficient in the production of radio jets, but are able to do so for both radiatively efficient (HERGs) and radiatively inefficient (LERGs) AGN, suggesting that it is not a correlation between the BH accretion state and jet power that drives the observational results in this section. \par

Another way in which the environment could influence the jet power under the B-P regime is through the jet lifetime. Although the halo mass does not directly influence the infall of the accretion material, a richer gas supply around the central BH in more massive haloes may lead to a prolonged continuous accretion process and, as a result, a \emph{longer lifetime} of the jet activity associated with it. Therefore, more energy is being deposited into the radio lobe, which is the main source of low-frequency radio emission from jet activity. However, if the enhanced quasar jet power in more massive haloes is tied to longer jet lifetimes, we would expect to see indications of order-of-magnitude older ages in the radio sources (i.e., larger radio sizes and more synchrotron-age spectra) residing in dense environments, which is not indicated in previous literature. Another way to test this assumption is to find robust evidence to whether the characteristic halo masses for quasar populations with different jet fractions are consistent across different cosmic epochs, which requires larger spectroscopic sample sets at higher redshifts.

\subsection{The role of black hole spin}

In the B-Z mechanism, the magnetic field is directly connected to the BH horizon, and the jet power scales roughly with the magnetic field strength $B$, the BH spin $a$, and the BH mass $M_\mathrm{BH}$ through $P\propto(BaM_\mathrm{BH})^2$ \citep[see e.g.][]{tchekhovskoy_efficient_2011,hardcastle_radio_2020}. Therefore, if the BH spin is correlated with the halo mass, then it can also drive a positive correlation between the halo mass and the jet power. \par

In common semi-analytical galaxy formation recipes, e.g. \textsc{galform} \citep{griffin_evolution_2019}, black holes can be spun up through two mechanisms. Firstly, the BH spin can increase with gas accretion, either through the starburst mode, where a cold gas supply from mergers is being transferred to the central BH engine through disc instability, or the hot halo mode where the collapse of massive haloes creates a shock in the hot gas, whose cooling time is longer than the free fall time and the gas therefore experiences a slow infall onto the accretion disc. Secondly, the BH spin can increase through dual-BH merger activities, where the final spin value depends on the spin of two merging BHs and the angular momentum of their orbit. Massive haloes can have a richer gas reservoir which enhances hot halo accretion, or the probability to have undergone a series of merger activities; both scenarios can help accelerate the spin-up of BHs, therefore boosting the radio jet power. \par

Previous results from numerical simulations have found that jet production is related to BHs spinning at maxima \citep[e.g.][]{benson_maximum_2009}, while the typical halo mass found in the hi-jet quasar population agrees with their model assumption. However, our observed correlation between halo mass and jet power will lead to a scenario where the BH spin scales with the halo mass, which spans over several orders of magnitude. Numerical models struggle to reproduce such a wide range of BH spin; even if they do, the simulated distributions of BH spin are peaked towards the high end, which contradicts the observations where only a small fraction of quasars host high-power jets \citep[e.g.][]{volonteri_evolution_2013}. This tension can be partly resolved by assuming an efficient spin-down paradigm during jet activities. During jet spin-down, the BH angular momentum is transferred away by radio jets, causing the BH spin to drop from the high initial values, and eventually broadening the BH spin distribution \citep[e.g.][]{husko_hybrid_2025}. Under this scenario, the more massive haloes are likely to have gone through a prolonged growth period which let the BHs to be spun up via accretion, and the haloes must continue to grow with cosmic time for powerful jets to remain active (or triggered) in later cosmic epochs, in order to balance with the jet spin-down. BHs in less massive haloes spin down faster, hence the lower jet power. The spin-down paradigm is therefore in agreement with the halo mass-jet power correlation, and also predicts a redshift evolutionary trend in the halo masses of jet-dominated quasars.

However, even when jet spin-down processes are included, only a factor of 2 difference between the upper and lower limits of BH spins is found at any given BH mass \citep{husko_hybrid_2025}. The correlation between BH spin and halo mass is also not trivial: one needs to assume all the BHs residing in rich environments to be spun-up while those in poor environments not to be, which lacks robust support from simulation results. We also would like to point out that the current observational data used to calibrate the numerical simulations on BH spin are heavily biassed toward the high-BH-spin end \citep[e.g.][]{reynolds_observational_2021}, therefore simulation models may not be able to reproduce the low-spin BHs by their nature. As a result, we conclude that BH spin is unlikely to be a major driver of the observed quasar radio jet power distribution; even if it does play a role, some other factor is likely to be needed as well.\par

\subsection{The importance of magnetic fields}

The other key factor that can potentially govern the jet power is the state of magnetic field close to the accreting black hole. \citet{tchekhovskoy_efficient_2011} and \citet{sikora_magnetic_2013} have developed a hybrid `magnetic flux' mechanism, in which the critical element to jet powering is the availability of a strong magnetic flux close to the BH, while the power of the magnetic flux is accumulated effectively under a hot accretion state (within a thick accretion disc) or long-lived accretion from a dense halo environment, eventually reaching a magnetic-arrested disc (MAD) state when the magnetic flux becomes saturated. However, this state does not \emph{guarantee} a jet; only when the accretion flow enters a state that supports strong magnetic stress, e.g., a cold (radiatively efficient) accretion phase with a magnitised corona (possibly through galaxy mergers), the system `switches on' and forms collimated jets. Under this scenario, the BHs may be threaded with a wide variety of magnetic fluxes due to the mix in the accretion disc state and merger progenitors, leading to the wide range of observed jet power. 

Previous studies including \citet{dunlop_quasars_2003} and \citet{matsuoka_massive_2014} suggest that quasars are typically hosted by massive galaxies and therefore should all have the potential to host hot accretion flows at some point in their past. The halo mass modulates the duty cycle and the power stored within the magnetic flux through the supply of accretion material, where the accretion processes become more extreme in denser environments due to cluster-scale cooling flow \citep[e.g.][]{fabian_observational_2012}, but does not impose a sharp threshold in the triggering of jets. This could naturally explain the continuous evolution and the lack of bimodality in the relation between halo mass and jet power. On the other hand, magnetic flux theory assumes that both radio quasars and LERGs require the presence of a hot accretion mode phase to accumulate the jet power; therefore, this framework can explain the similarity between HERGs and LERGs halo environments, as the only difference between the two populations is whether the AGN is \emph{currently} under a hot accretion state (LERGs) or not (HERGs). Since the launch of magnetic flux-driven jet does not require an \emph{active} hot accretion state, the theory also accounts for the lack of correlation between quasar jet power and BH mass/Eddington ratio. \par

Although we cannot definitively rule out any of the possible mechanisms discussed above due to our lack of knowledge in the halo/BH merger history and poor observational constraints on the BH spin and magnetic field strength, the magnetic flux model, possibly in combination with variations in BH spin, currently has all the ingredients to link the quasar jet power and environment together.

 \section{Summary}
\label{sec:conclusion}

In this work, building on the two-component Bayesian model presented in \citet{yue_novel_2024}, we established a new definition of radio quasar populations motivated by physical processes rather than single thresholds of radio luminosity or radio loudness. We classified the eBOSS quasars into four populations based on their dominant source of radio emission: host galaxy star formation dominated (SFQ), radio intermediate (RI), low-power AGN jet dominated (lo-jet), and high-power AGN jet dominated (hi-jet). We then investigated the spatial clustering of these radio quasar populations by measuring auto-correlation TPCFs in each population at different redshift slices. Our results reveal a positive correlation between the correlation lengths and the fraction of jet activity in radio emission, where quasars become increasingly clustered as the radio jet power increases from SFQs to hi-jet quasars. Indicative estimations on the quasar bias and halo mass show that the characteristic mass of the dark matter haloes hosting each quasar population also increases with the jet fraction, where SFQs reside in similar haloes with the full SDSS quasar sample ($M\sim10^{12.5}h^{-1}\ M_\odot$), while jet-dominated quasars reside in rich cluster environments with $M_\mathrm{halo}\sim10^{14}h^{-1}\ M_\odot$. \par

We show that the BH mass difference only has a minor influence in the halo mass difference between radio quasar populations, and bolometric luminosity does not affect the average halo mass in all populations. The relationship between halo mass and jet power is therefore fundamental and not a secondary effect due to other factors correlating with the halo mass and jet power. Our result rules out the bimodality in large-scale environments between the traditionally defined RL and RQ quasars, suggesting that there is no minimum halo mass or BH mass required for the triggering of radio jets. The continuous evolution between halo mass and jet power into the jet-dominated regime suggests that the halo mass is responsible for setting the power of radio jets. \par

Our results pose new constraints on the current AGN jet production models. As we do not observe major difference between the halo masses of jet-dominated quasars (selected in optical bands, mostly HERGs) and the radio AGN (selected in radio bands, mostly LERGs), we speculate that the halo environment has little influence on determining the BH accretion modes when radio jets are present, and therefore AGN in hot and cold accretion modes share similar triggering and powering mechanisms of radio jets. Spin-powered jets may contribute to the observed halo-jet correlation through the spin-down mechanism; however, this mechanism cannot fully explain the wide distribution of jet powers and is unlikely to play a major role in establishing the correlation. The `magnetic flux' model, where radio jets are triggered by a state that supports strong magnetic stress (e.g., a cold accretion phase with a magnetised corona), with power determined by the magnetic flux accumulated during the hot accretion phase (or through long-lived accretion from a dense halo environment), can account for all of the observed correlations in this work, and is more likely the main physical driver behind the production of AGN radio jets. \par

Future observational campaigns will open the window for more careful examinations on different jet production theories through clustering analysis. Since low-frequency radio bands are more sensitive in older electrons compared with higher-frequency radio observations, complementary results from 1.4 GHz (or higher frequency) radio bands with similar sensitivity and spatial coverage may provide constraints on the radio spectral slope and lifetime of observed jet activities, which would help test for the prolonged accretion scenario. More complete X-ray samples from future wide-field surveys including the SRG/eROSITA all-sky survey \citep[eRASS;][]{merloni_srgerosita_2024} will improve the statistics of X-ray-detected galaxy groups and clusters, which could directly verify the link between rich environments and jet incidence. High surface brightness-sensitivity images from the \emph{Euclid} satellite would greatly expand the current sample selection of galaxies and quasars with identified merger processes \citep{euclid_collaboration_euclid_2025}, allowing statistical studies on the direct impact of galaxy mergers on jet production. Finally, improved statistics from next-generation spectroscopical surveys, including the Dark Energy Spectroscopic Instrument experiment \citep[DESI;][]{desi_collaboration_desi_2016,chaussidon_target_2023} and SDSS-V, together with the additional radio sky coverage of LoTSS DR3, will enable detailed studies on the halo assembly history of different radio quasar populations through HOD modelling, allowing us to test for the probability of jets being triggered at different halo masses and cosmic epochs.

\section*{Acknowledgements}

The authors would like to thank John Peacock, Dave Alexander, and Chris Done for the useful discussions and comments that helped to shape this paper. BY acknowledges support from the University of Edinburgh and Leiden Observatory through the Edinburgh-Leiden joint studentship. KJD acknowledges funding from the STFC through an Ernest Rutherford Fellowship (grant number ST/W003120/1). CLH acknowledges support from the Oxford Hintze Centre for Astrophysical Surveys which is funded through generous support from the Hintze Family Charitable Foundation. LKM is grateful for support from a UKRI FLF [MR/Y020405/1]. Several authors are grateful for the support of the UK Science and Technology Facilities Council (STFC) through grants ST/Y000951/1 (PNB\&CLH), ST/V000624/1 (DJBS), and ST/Y001028/1 (DJBS). \par


LOFAR data products were provided by the LOFAR Surveys Key Science project (LSKSP; \url{https://lofar-surveys.org/}) and were derived from observations with the International LOFAR Telescope (ILT). LOFAR \citep[][]{van_haarlem_lofar_2013}{}{} is the Low Frequency Array designed and constructed by ASTRON. It has observing, data processing, and data storage facilities in several countries, which are owned by various parties (each with their own funding sources), and which are collectively operated by the ILT foundation under a joint scientific policy. The efforts of the LSKSP have benefited from funding from the European Research Council, NOVA, NWO, CNRS-INSU, the SURF Co-operative, the UK Science and Technology Funding Council and the J\"ulich Supercomputing Centre. This research made use of the University of Hertfordshire high-performance computing facility and the LOFAR-UK computing facility located at the University of Hertfordshire and supported by STFC [ST/P000096/1].\par

Funding for the Sloan Digital Sky Survey IV has been provided by the Alfred P. Sloan Foundation, the U.S. Department of Energy Office of Science, and the Participating Institutions. \par

SDSS-IV acknowledges support and resources from the Center for High Performance Computing at the University of Utah. The SDSS website is \url{www.sdss4.org}. \par

SDSS-IV is managed by the Astrophysical Research Consortium for the Participating Institutions of the SDSS Collaboration including the Brazilian Participation Group, the Carnegie Institution for Science, Carnegie Mellon University, Center for Astrophysics | Harvard \& Smithsonian, the Chilean Participation Group, the French Participation Group, Instituto de Astrof\'isica de Canarias, The Johns Hopkins University, Kavli Institute for the Physics and Mathematics of the Universe (IPMU) / University of Tokyo, the Korean Participation Group, Lawrence Berkeley National Laboratory, Leibniz Institut f\"ur Astrophysik Potsdam (AIP),  Max-Planck-Institut f\"ur Astronomie (MPIA Heidelberg), Max-Planck-Institut f\"ur Astrophysik (MPA Garching), Max-Planck-Institut f\"ur Extraterrestrische Physik (MPE), National Astronomical Observatories of China, New Mexico State University, New York University, University of Notre Dame, Observat\'ario Nacional / MCTI, The Ohio State University, Pennsylvania State University, Shanghai Astronomical Observatory, United Kingdom Participation Group, Universidad Nacional Aut\'onoma de M\'exico, University of Arizona, University of Colorado Boulder, University of Oxford, University of Portsmouth, University of Utah, University of Virginia, University of Washington, University of Wisconsin, Vanderbilt University, and Yale University.

\section*{Data Availability}

The datasets used in this paper were derived from sources in the public domain: the LOFAR Two-Metre Sky Surveys (\url{www.lofar-surveys.org}) and the Sloan Digital Sky Survey (\url{www.sdss.org}).



\bibliographystyle{mnras}
\bibliography{references} 




\clearpage
\appendix
\section{Fitted values from two-point correlation functions for various quasar sub-samples }

In Table~\ref{tab:result_1} and \ref{tab:result_2} we present the fitted redshift space correlation lengths ($s_0/h^{-1}\ \mathrm{Mpc}$) and linear bias ($b$) from the two-point correlation functions for the various quasar sub-samples defined in this work, including SFQs, RIQs, lo-jet and hi-jet quasars, and quasars split by BH mass and \emph{i}-band luminosities. All values are fitted with a single power law with $\xi_s(s)=(s/s_0)^{-\gamma}$, where $\xi_s(s)$ is the two-point correlation function and $\gamma$ is fixed at $\gamma=1.95$ based on the result of the entire quasar sample, except for the full quasar samples where the best-fit values of $\gamma$ are listed explicitly (Figure~\ref{fig:clustering_amp_full}). We also list the number of quasars used in each fit ($N_\textrm{QSO}$).

\clearpage

\begin{landscape}

\begin{table}
    \centering
    \begin{tabular}{llccc>{\centering\arraybackslash}p{0.08\linewidth}ccc}
        \toprule
         All quasars & & $0.8<z<1.0$ & $1.0<z<1.2$ & $1.2<z<1.4$ & $1.4<z<1.6$ & $1.6<z<1.8$ & $1.8<z<2.0$ & $2.0<z<2.2$ \\
         & $N_\textrm{QSO}$& 19616& 25518& 30759& 30082& 30227& 27104&21538\\
         & $\gamma$& $1.96\pm0.09$& $1.90\pm0.07$& $2.01\pm0.09$& $1.88\pm0.10$& $1.96\pm0.10$& $2.01\pm0.12$&$1.95\pm0.13$\\
         & $s_0\ [h^{-1}\ \mathrm{Mpc}]$ &  $7.12\pm0.28$&  $6.88\pm0.25$&  $7.28\pm0.28$&  $7.16\pm0.34$&  $7.24\pm0.32$&  $7.63\pm0.41$& $7.95\pm0.47$\\
         &  &  &  &  &  &  &  & \\
         SF-dominated (SFQs) & & $0.8<z<1.0$ & $1.0<z<1.2$ & $1.2<z<1.4$ & $1.4<z<1.6$ & $1.6<z<1.8$ & $1.8<z<2.0$ & $2.0<z<2.2$ \\
         & $N_\textrm{QSO}$& 15817& 20883& 27475& 25723& 25229& 22131&17559\\
         & $s_0\ [h^{-1}\ \mathrm{Mpc}]$ &  $7.17\pm0.25$&  $7.08\pm0.22$&  $7.03\pm0.22$&  $7.34\pm0.24$&  $7.16\pm0.23$&  $7.26\pm0.26$& $7.58\pm0.43$\\
         & $b_\mathrm{QSO}$ &  $1.73\pm0.06$&  $1.86\pm0.06$&  $2.01\pm0.06$&  $2.26\pm0.07$&  $2.37\pm0.08$&  $2.57\pm0.09$& $2.85\pm0.16$\\
         &  &  &  &  &  &  &  & \\
         Radio intermediate (RIQs) & & $0.8<z<1.0$ & $1.0<z<1.2$ & $1.2<z<1.4$ & $1.4<z<1.6$ & $1.6<z<1.8$ & $1.8<z<2.0$ & $2.0<z<2.2$ \\
         & $N_\textrm{QSO}$& 928& 1808& 1774& 2474& 2092& 1834&1296\\
         & $s_0\ [h^{-1}\ \mathrm{Mpc}]$ &  $8.82\pm1.41$&  $9.17\pm1.92$&  $8.79\pm1.66$&  $9.41\pm1.79$&  $9.44\pm1.10$&  $9.30\pm1.50$& $9.57\pm1.11$\\
         & $b_\mathrm{QSO}$ &  $2.11\pm0.33$&  $2.39\pm0.49$&  $2.49\pm0.46$&  $2.88\pm0.54$&  $3.10\pm0.35$&  $3.27\pm0.51$& $3.58\pm0.41$\\
         &  &  &  &  &  &  &  & \\
         low-power jet dominated (lo-jet) & & $0.8<z<1.4$& $1.4<z<1.8$& $1.8<z<2.2$& & & & \\
         & $N_\textrm{QSO}$& 1566& 1546& 1402& & & &\\
         & $s_0\ [h^{-1}\ \mathrm{Mpc}]$ &  $11.41\pm2.49$&  $11.85\pm1.62$&  $11.77\pm2.28$&  &  &  & \\
         & $b_\mathrm{QSO}$ &  $2.96\pm0.63$&  $3.74\pm0.50$&  $4.25\pm0.80$&  &  &  & \\
         &  &  &  &  &  &  &  & \\
         high-power jet dominated (hi-jet) & & $0.8<z<1.4$& $1.4<z<1.8$& $1.8<z<2.2$& & & & \\
         & $N_\textrm{QSO}$& 1092& 927& 753& & & &\\
         & $s_0\ [h^{-1}\ \mathrm{Mpc}]$ &  $16.40\pm1.52$&  $15.74\pm2.10$&  $15.95\pm2.78$&  &  &  & \\
         & $b_\mathrm{QSO}$ &  $4.22\pm0.38$&  $4.93\pm0.64$&  $5.72\pm0.97$&  &  &  & \\
         \bottomrule
    \end{tabular}
    \caption{Tabulated values of the correlation length ($s_0$) and the derived linear bias ($b$; only for quasars split by jet fraction) fitted from the TPCFs of different quasar sub-populations defined in this work, measured in various redshift slices. The number of quasars used in each fit is listed in rows of $N_\textrm{QSO}$.}
    \label{tab:result_1}
\end{table}
\end{landscape}

\begin{landscape}
\begin{table}
    \centering
    \begin{tabular}{llccc>{\centering\arraybackslash}p{0.08\linewidth}ccc}
        \toprule
         $0-80\%\ M_\mathrm{BH}$, SF-dominated & & $0.8<z<1.0$ & $1.0<z<1.2$ & $1.2<z<1.4$ & $1.4<z<1.6$ & $1.6<z<1.8$ & $1.8<z<2.0$ & $2.0<z<2.2$ \\
         & $N_\textrm{QSO}$& 13394& 16390& 22550& 20523& 20748& 17453&14063\\
         & $s_0\ [h^{-1}\ \mathrm{Mpc}]$ &  $6.98\pm0.28$&  $6.87\pm0.26$&  $6.96\pm0.19$&  $7.46\pm0.20$&  $7.16\pm0.26$&  $7.33\pm0.26$& $7.69\pm0.42$\\
         &  &  &  &  &  &  &  & \\
         $80\%-100\%\ M_\mathrm{BH}$, SF-dominated & & $0.8<z<1.0$ & $1.0<z<1.2$ & $1.2<z<1.4$ & $1.4<z<1.6$ & $1.6<z<1.8$ & $1.8<z<2.0$ & $2.0<z<2.2$ \\
         & $N_\textrm{QSO}$& 2418& 4472& 4867& 5153& 4403& 4596&3467\\
         & $s_0\ [h^{-1}\ \mathrm{Mpc}]$ &  $7.77\pm1.19$&  $8.12\pm0.54$&  $7.94\pm0.68$&  $7.78\pm0.40$&  $8.31\pm0.57$&  $7.74\pm0.87$& $9.71\pm0.98$\\
         &  &  &  &  &  &  &  & \\
         $0-80\%\ M_\mathrm{BH}$, jet-dominated & & $0.8<z<1.6$& $1.6<z<2.2$& & & & & \\
         & $N_\textrm{QSO}$& 2546& 2424& & & & &\\
         & $s_0\ [h^{-1}\ \mathrm{Mpc}]$ &  $11.47\pm3.28$&  $12.02\pm1.04$& & & & & \\
         $80-100\%\ M_\mathrm{BH}$, jet-dominated& & $0.8<z<1.6$& $1.6<z<2.2$& & & & & \\
         & $N_\textrm{QSO}$& 1416& 879& & & & &\\
         & $s_0\ [h^{-1}\ \mathrm{Mpc}]$ &  $13.42\pm7.88$&  $17.30\pm3.47$&  &  &  &  & \\
         \midrule
         Bright (top 50\% \emph{i}-band luminosity) & & $0.8<z<1.0$ & $1.0<z<1.2$ & $1.2<z<1.4$ & $1.4<z<1.6$ & $1.6<z<1.8$ & $1.8<z<2.0$ & $2.0<z<2.2$ \\
         & $N_\textrm{QSO}$& 8165& 12314& 14059& 15874& 13336& 13464&9644\\
         & $s_0\ [h^{-1}\ \mathrm{Mpc}]$ &  $7.96\pm0.48$&  $7.40\pm0.19$&  $7.70\pm0.38$&  $7.34\pm0.25$&  $7.41\pm0.35$&  $7.96\pm0.28$& $7.87\pm0.47$\\
         &  &  &  &  &  &  &  & \\
         Faint (bottom 50\% \emph{i}-band luminosity) & & $0.8<z<1.0$ & $1.0<z<1.2$ & $1.2<z<1.4$ & $1.4<z<1.6$ & $1.6<z<1.8$ & $1.8<z<2.0$ & $2.0<z<2.2$ \\
         & $N_\textrm{QSO}$& 9144& 11321& 16340& 13636& 14887& 11977&10148\\
         & $s_0\ [h^{-1}\ \mathrm{Mpc}]$ &  $7.20\pm0.32$&  $7.87\pm0.35$&  $7.06\pm0.36$&  $7.94\pm0.35$&  $7.25\pm0.23$&  $7.02\pm0.32$& $7.82\pm0.62$\\
         &  &  &  &  &  &  &  & \\
         $10\%$ most luminous & & $0.8<z<1.0$ & $1.0<z<1.2$ & $1.2<z<1.4$ & $1.4<z<1.6$ & $1.6<z<1.8$ & $1.8<z<2.0$ & $2.0<z<2.2$ \\
         & $N_\textrm{QSO}$& 1512& 2584& 2741& 3237& 2604& 2714&1879\\
         & $s_0\ [h^{-1}\ \mathrm{Mpc}]$ &  $7.64\pm1.24$&  $7.22\pm1.11$&  $7.52\pm1.22$&  $8.41\pm0.80$&  $7.94\pm0.89$&  $7.02\pm0.75$& $9.64\pm1.48$\\
         &  &  &  &  &  &  &  & \\
         $90\%$ least luminous & & $0.8<z<1.0$ & $1.0<z<1.2$ & $1.2<z<1.4$ & $1.4<z<1.6$ & $1.6<z<1.8$ & $1.8<z<2.0$ & $2.0<z<2.2$ \\
         & $N_\textrm{QSO}$& 15797& 21051& 27658& 26273& 25619& 22727&17913\\
         & $s_0\ [h^{-1}\ \mathrm{Mpc}]$ &  $7.31\pm0.25$&  $7.16\pm0.21$&  $7.37\pm0.24$&  $7.64\pm0.23$&  $7.16\pm0.22$&  $7.44\pm0.24$& $7.53\pm0.42$\\
         \bottomrule
    \end{tabular}
    \caption{(Continued from Table~\ref{tab:result_1}.)}
    \label{tab:result_2}
\end{table}

\label{lastpage}
\end{landscape}

\end{document}